\documentclass[prb,preprint,showpacs]{revtex4}
\usepackage{graphicx}
\usepackage{color}
\DeclareGraphicsExtensions{pdf,jpg}
\usepackage{amsmath,amssymb,bm} 

\begin{document}
\title{Measuring strain and rotation fields at the dislocation core in graphene }
\author{L.L. Bonilla$^{1,2}$, A. Carpio$^{3}$, C. Gong$^4$, J.H. Warner$^4$ }
\affiliation {
$^1$G. Mill\'an Institute, Fluid Dynamics, Nanoscience and Industrial
Mathematics, Universidad Carlos III de Madrid, Avda.\ Universidad 30; E-28911 Legan\'es, Spain\\
$^2$Unidad Asociada al Instituto de Ciencia de Materiales de Madrid, CSIC, 28049 Cantoblanco, Madrid, Spain\\
$^3$Departmento de Matem\'atica Aplicada, Universidad Complutense de Madrid; E-28040 Madrid, Spain\\
$^4$Department of Materials, University of Oxford, Parks Road,
Oxford OX1 3PH, UK. \\ 
Corresponding author: bonilla@ing.uc3m.es (LLB)}

\renewcommand{\thefootnote}{\arabic{footnote}}

\begin{abstract}
Strain fields, dislocations and defects may be used to control electronic properties of graphene. By using advanced imaging techniques with high-resolution transmission electron microscopes, we have measured the strain and rotation fields about dislocations in monolayer graphene with single-atom sensitivity. These fields differ qualitatively from those given by conventional linear elasticity. However, atom positions calculated from two dimensional (2D) discrete elasticity and three dimensional discrete periodized F\"oppl-von K\'arm\'an equations (dpFvKEs) yield fields close to experiments when determined by geometric phase analysis. 2D theories produce symmetric fields whereas those from experiments exhibit asymmetries. Numerical solutions of dpFvKEs provide strain and rotation fields of dislocation dipoles and pairs that also exhibit asymmetries and, compared with experiments, may yield information on out-of-plane displacements of atoms. While discrete theories need to be solved numerically, analytical formulas for strains and rotation about dislocations can be obtained from 2D Mindlin's hyperstress theory. These formulas are very useful for fitting experimental data and provide a template to ascertain the importance of nonlinear and nonplanar effects. Measuring the parameters of this theory, we find two characteristic lengths between three and four times the lattice spacings that control dilatation and rotation about a dislocation. At larger distances from the dislocation core, the elastic fields decay to those of conventional elasticity. Our results may be relevant for strain engineering in graphene and other 2D materials of current interest.
\end{abstract}
\maketitle

\section{Introduction} \label{sec:1}
Strain fields, dislocations and defects may be used to control electronic properties of graphene\cite{gui10,lev10,nai12,gun11}. Defects play a key role in graphene physicochemical properties and could be critical to generate biologically compatible materials and sensors \cite{ter12}. Experiments have shown that vacancies and point defects produce paramagnetism\cite{nai12} and appropriate strain fields induce strong pseudo-magnetic fields and Landau levels\cite{lev10}. This may pave the way to extending graphene electronic applications into spin-based technology. Special line defects (grain boundaries) may be used to filter electrons from different graphene valleys (valleytronic devices)\cite{gun11}. Dislocations play an important role in finding an accurate description of both elasticity and plasticity in graphene. Dislocations deform graphene by elongation and compression of C-C bonds, shear, and lattice rotations as shown\cite{war12} by advanced imaging techniques with high-resolution transmission electron microscopes (HRTEMs)\cite{hyt03,zha08}. In graphene and other two-dimensional (2D) crystals, images of dislocation cores as defects in the crystal lattice and real-time pictures of defect evolution are now being obtained\cite{mey08,mao11}. Dislocation motion and defect groupings may be accompanied by markedly long-ranging out-of-plane buckling \cite{leh13,kot14} (theory in Refs. \onlinecite{SN88,CN93}) and facilitated by accommodating extra carbon atom pairs in out-of-plane blisters formed by several dislocation pairs \cite{rob14}. Recent experiments show that direct implantation of carbon atoms in graphene creates interstitial aggregates and dislocation dipoles that disappear at larger HRTEM electron doses.\cite{leh15} Dislocation pairs with dislocations having Burgers vectors pointing to each other\cite{leh13} and dipoles whose dislocations have Burgers vectors pointing away from each other\cite{lee14} display out-of-plane buckling. At the micrometer scale, there are effective computational theories of line dislocations that rely on a mixture of theory and empirical rules for dislocation interaction and motion\cite{bul06}. Precise measurements and theoretical understanding at the nanometer scale could help bridging the gap between scales.

In this paper, we have measured the strain and rotation fields about dislocations in monolayer graphene with single-atom sensitivity. Our measures suggest that the observed strong discrepancies of the strain and rotation fields of graphene with conventional elasticity in the presence of defects are due to {\em nonlinear lattice effects} that are most noticeable in the immediate neighborhood of defects but decay in the far field thereof. While experimentally imaged areas often contain multiple defects and distortions, dislocations that are well separated from other defects present in the lattice allow more definite tests of theories. Since the elastic fields of dislocations in graphene decay in about four lattice spacings to those given by continuum elasticity,\cite{zha06} well separated dislocations present smaller regions in which continuum elasticity needs to be modified. We show that strain and rotation fields determined from experiments by using Geometric Phase Analysis (GPA)\cite{hyt03,hyt98,hrem} may be compared to results of planar and nonplanar theoretical approaches with increasing agreement. The comparison permits to infer the in-plane and out-of-plane displacements of the carbon atoms near dislocations in the graphene sheet. 

Firstly, we can place atoms on a planar hexagonal lattice using the well known 2D linear elasticity formula for the displacement vector about a dislocation \cite{ll7} if the dislocation point is appropriately selected. When we treat these positions using GPA, we obtain strain and rotation fields that are {\em qualitatively} similar to those revealed  by experiments for dislocations  having one the six possible basic Burgers vectors along primitive directions. More refined 2D discrete elasticity,\cite{cb05} periodized along the primitive directions of the hexagonal lattice,\cite{car08,CBJV08,bon11} produces more accurate strains and rotation that are compatible with observations. Furthermore, discrete periodized theories capture pinning and depinning of defects by the lattice.\cite{car01,CB03} We also show that higher order 2D elasticity theories give useful {\em explicit formulas} for strains and rotation that can be calibrated to produce good qualitative and quantitative (up to 10\% errors) agreement with experiments. In general, 2D theories tend to produce symmetric strain fields about dislocations whereas strains obtained from experiments exhibit asymmetries. Nonplanar effects inducing asymmetries are captured by three-dimensional (3D) F\"oppl-von K\'arm\'an equations (FvKEs) discretized on the hexagonal lattice and periodized along primitive directions.\cite{bon12,bonJSTAT12} For dipoles and dislocation pairs comprising well-separated dislocations, discrete periodized FvKEs (dpFvKEs) predict four stable configurations for which the graphene sheet is {\em buckled}. In two configurations, the sheet is asymmetrically buckled, with one dislocation on a hill and the other in a trough. In the other two configurations, both dislocations sit either on a hill or on a trough. Similar results have been reported for dislocation loops using energy methods,\cite{leh13} and for dislocation dipoles using density functional theory (DFT),\cite{lee14} but not dynamic simulations of the dpFvKEs as in the present paper. By comparing GPA of experiments and of predictions by the discrete periodized F\"oppl-von K\'arm\'an equations, we can select three dimensional configurations that provide better agreement. This strategy that infers out-of-plane displacements from the experimental image may be limited by the combined effect of neglecting defects present in the image and of errors introduced by the numerical procedure and the image processing software.

In short, GPA based on atom positions calculated by means of the 2D displacement vector of linear  elasticity gives a fair prediction of the observed strain and rotation fields around isolated defects in graphene, whereas strains and rotation found by differentiating the formula for the displacement vector (i.e., the strains and rotation given by conventional elasticity) are way off. To get a better quantitative agreement with experiments, we need to regularize continuum 2D planar elasticity and 3D F\"oppl-von K\'arm\'an equations on the graphene lattice. An alternative to 2D discrete elasticity that gives analytical expressions for strains and rotation is hyperstress theory. These expressions are very useful for fitting experimental data and provide a template to ascertain the importance of nonplanar effects. That a continuum theory (linear elasticity based on displacement vectors) gives a fair approximation to elastic fields at the \AA ngstrom scale of dislocation cores is quite unexpected. The agreement with experiments improves when we consider nonlinear lattice and nonplanar effects by using 3D discrete periodized F\"oppl-von K\'arm\'an equations. The combination of planar and nonplanar theories and experiments allows to infer the more likely atomic configuration in dislocation cores.

The rest of the paper is as follows. Section \ref{sec:2} contains the results of experiments and numerical simulations of different theories used to interpret them. The detailed methodology used to obtain the theoretical predictions is explained in the other sections. Section \ref{sec:GPA} describes how the Geometrical Phase Analysis method produces strain and rotation fields from images of atoms (that may come from experimental images or from simulation results). Section \ref{sec:3} describes planar and nonplanar discrete elasticity arising from discretizing 2D linear elasticity and 3D F\"oppl-von K\'arm\'an equations on the hexagonal lattice and replacing finite differences along primitive directions by periodic functions thereof. Section \ref{sec:4} presents 2D Mindlin hyperstress elasticity \cite{min64}, and it explains how parameters of the theory are fitted with data from experiments. Section \ref{sec:5} contains a guide to the results from different theories and it explains how to obtain the strains, rotation and errors in the different figures. The last section summarizes our conclusions. Appendix \ref{app} contains the derivation of the displacement vector of a dislocation in the hyperstress framework.

\section{Results and interpretation} \label{sec:2}

\begin{figure}
\includegraphics[width=9cm]{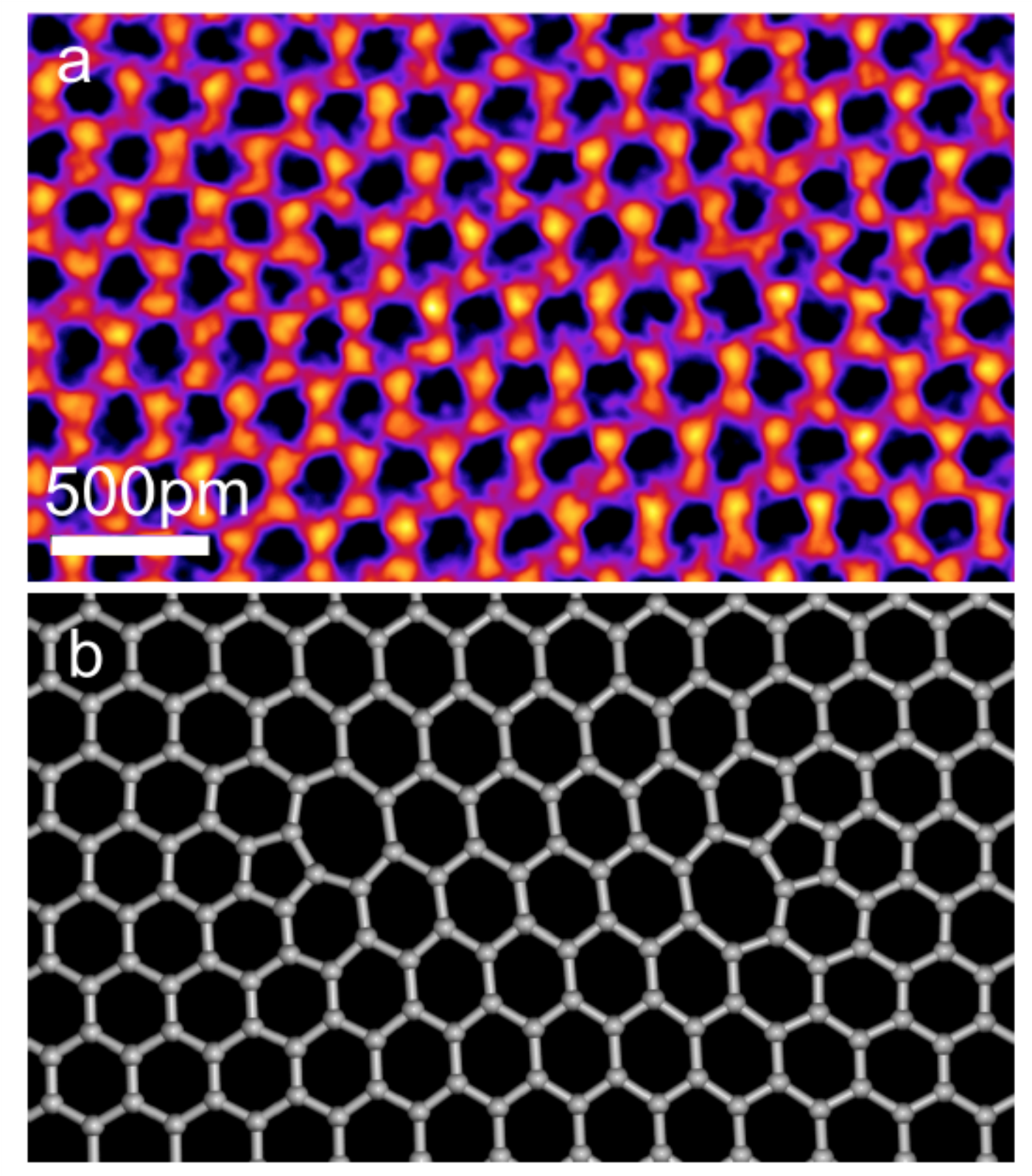}
\caption{(Color online) (a) HRTEM image of a dislocation dipole in graphene formed by controlled electron beam irradiation at an accelerating voltage of 80 kV. (b) Atomic model representing the dislocation dipole shown in (a). Their Burgers vectors of the component dislocations (separated by 4 atomic spacings) are $\pm(1/2,-\sqrt{3}/2)$ and the lines of extra atoms of the dislocations point away from each other. We measure all lengths in units of the lattice spacing $a = 2.46$ \AA.}
\label{fig1}
\end{figure}

Recent work has shown that controlled focused electron beam irradiation can lead to the formation of dislocation dipoles in graphene due to sputtering of atoms along the zig-zag lattice direction. This method starts with a pristine region of graphene, followed by the introduction of the defects by electron beam irradiation, and then the atomic resolution study of the graphene lattice distortion by HRTEM. By maintaining the sample within the vacuum chamber of the HRTEM it reduces the probability of other atoms attaching to the defects and modifying the structure and strain in the system. Figure \ref{fig1}(a) shows a HRTEM image of a dislocation dipole formed in graphene, with an atomic model in Figure \ref{fig1}(b). The study of the dynamics of dislocations has shown they are mobile and can exhibit gliding and climbing motion, likely driven by energy supplied by the electron beam\cite{war12}. GPA produces the strain and rotation fields near dislocations \cite{war12,leh13,rob14}. Extended observations over long times allow to identify extra pairs of atoms and blister defects in the lattice that help dislocation motion\cite{rob14}. 

\begin{figure}
\includegraphics[width=15cm]{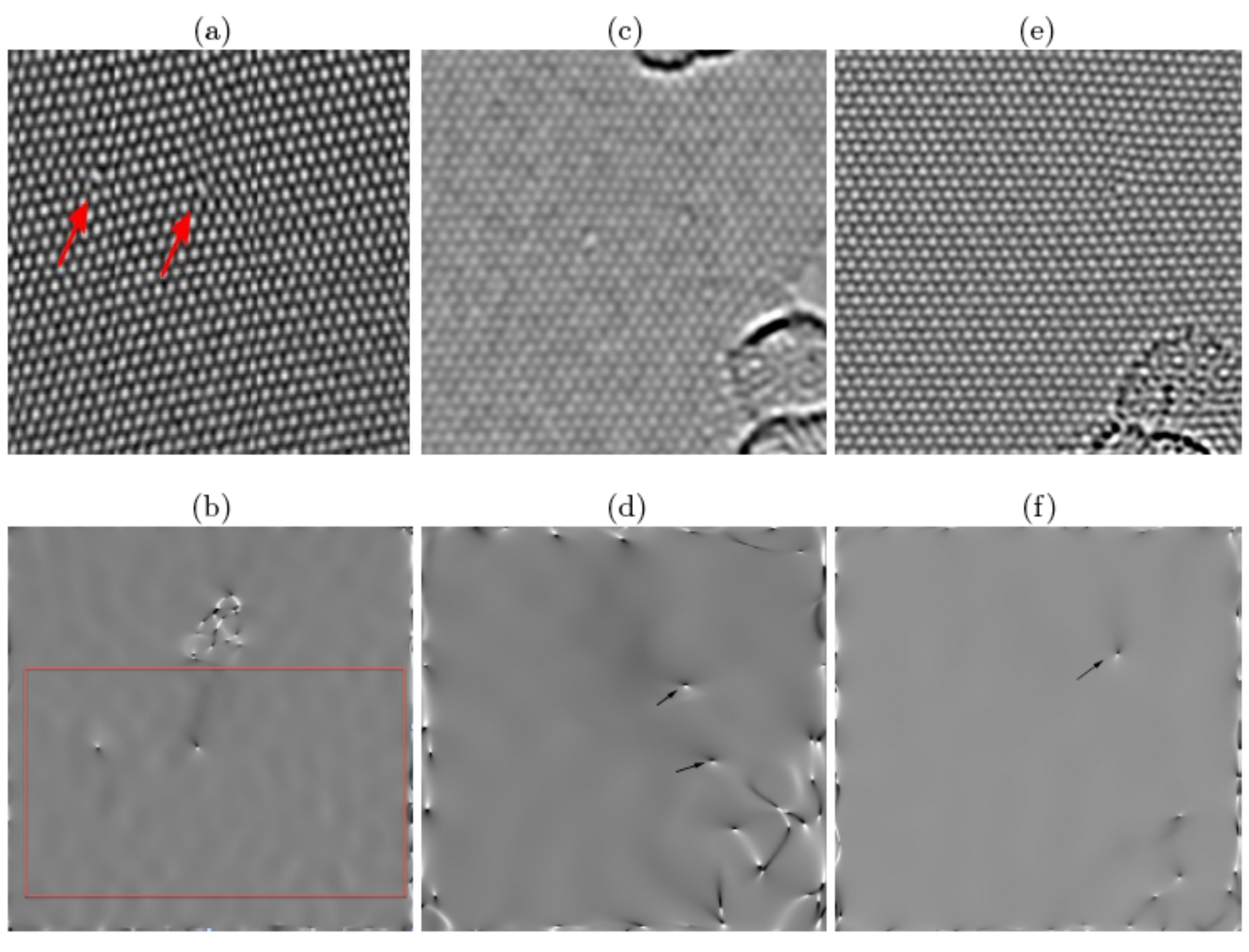}
\caption{(a) Dislocation dipole with Burgers vectors $(-1,\sqrt{3})/2$ (left dislocation) and $(1,-\sqrt{3})/2$ (right dislocation), similar to that in Figure \ref{fig1} but now the dislocations are separated by 8 lattice spacings. (b) GPA generated strain field $e_{xx}$ of a larger region that includes (a) inside the marked rectangle. Besides the dislocations, there is a cluster of neighboring defects. (c) Two dislocations with Burgers vectors $(1,0)$ separated from regions with many defects. (d) Strain field $e_{xx}$ corresponding to (c) generated by GPA. Besides the dislocations, marked by arrows, we observe grain boundaries. (e) Isolated dislocation with Burgers vector $(1,\sqrt{3})/2$. (f) Strain field $e_{xx}$ corresponding to (e) generated by GPA.  In (b), (d) and (f) there are spurious defects at the lattice border due to the finite size of the image. The presence of grain boundaries, ripples, holes and other defects in the far field may distort the strain and rotation fields of the dislocations. The noise introduced by true and spurious defects is reduced increasing the distance to the dislocation under study. }
\label{fig2}
\end{figure}

In our experiments, graphene was produced by chemical vapor deposition on a liquid copper catalyst according to a previously reported method\cite{wu12}. The graphene surface was spin coated with a solution containing poly methyl methacrylate (PMMA) to provide a support for transfer. The Cu was etched away using FeCl$_3$, and the graphene: PMMA sample was transferred into HCl, followed by several water rinses. After transferring to holey silicon nitride HRTEM grids, the sample was baked in air at 300 C overnight. Atomic resolution images of dislocations in graphene were obtained using the Oxford-JEOL 2200MCO HRTEM, with CEOS probe and image correctors, operating at 80 kV. Monochromation of the electron beam was performed to give an energy spread of $\sim$0.3 meV. This provides resolution sufficient to resolve the position of atoms in graphene. 

We have analyzed the images of a number of dislocation configurations with Burgers vectors $\pm (1,0)$, $\pm (1/2,-\sqrt{3}/2)$ and $\pm (1/2,\sqrt{3}/2)$ (all lengths are measured in units of the lattice spacing $a$). Some are shown in Figure \ref{fig2}. GPA of experimental images (cf.Ê Section~\ref{sec:GPA}) produces the rotation and strains of the dislocations that are then compared to theoretical predictions (cf.Ê Sections~\ref{sec:3}-\ref{sec:5}). In all cases, 2D predictions obtained by GPA of lattices deformed according to off-centered  elastic dislocation displacement fields or to discrete periodized planar elasticity solutions, and 2D predictions obtained by hyperstress fittings, provide a good agreement with experimental observations. In specific situations, we have compared the experimental strain and rotation fields to the GPA of lattices representing pair or dipole solutions of 3D discrete periodized F\"oppl-Von K\'arm\'an equations, finding one configuration that seemed to furnish the best fit, up to errors due to numerical approximations, image processing or the presence of neglected defects. This strategy may be useful to distinguish out-of-plane effects, depending on the limiting circumstances. Figure \ref{fig2} selects some significative tests that we now discuss in more detail.

\begin{figure}
\centering
\includegraphics[width=15cm]{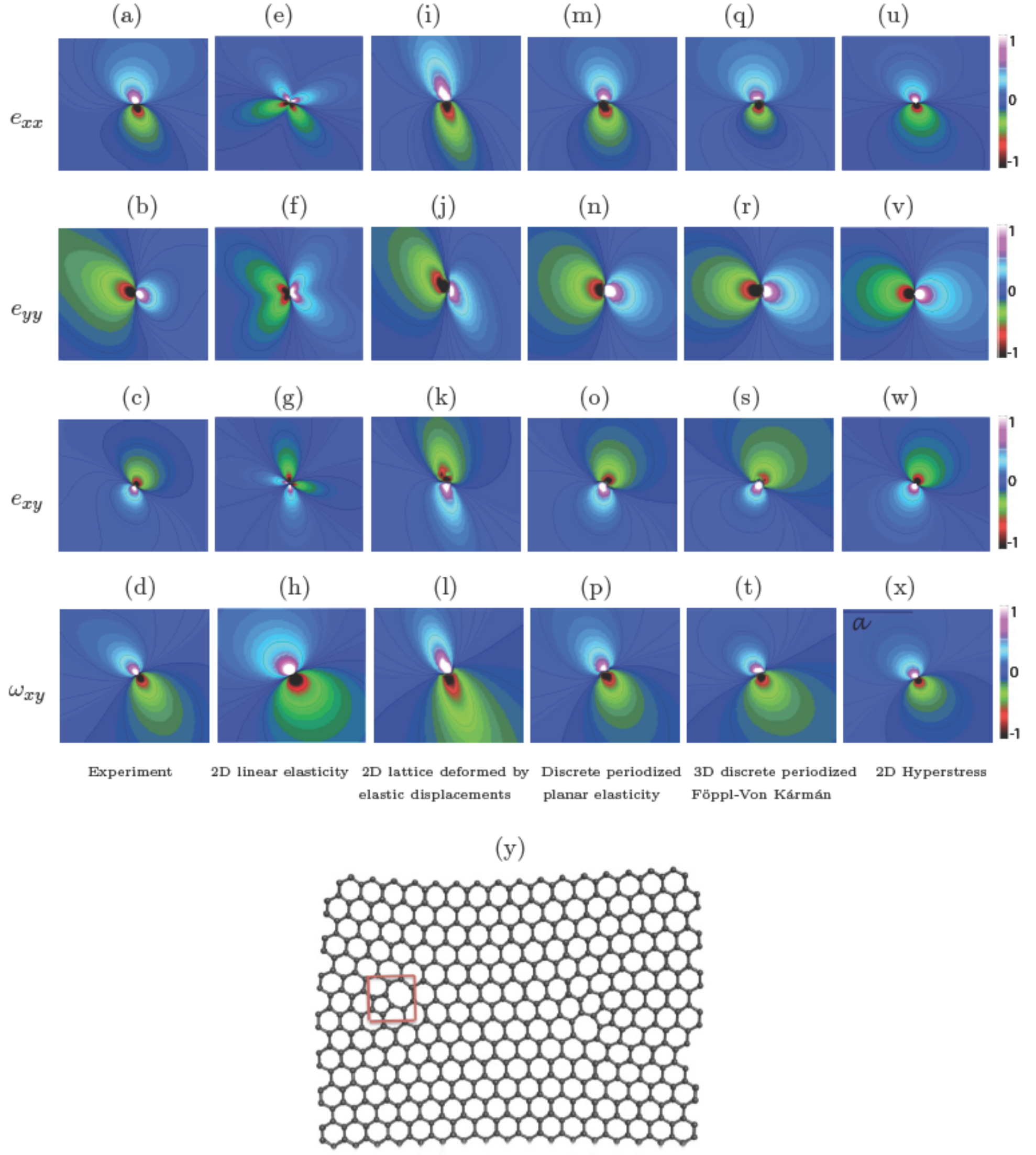} 
\caption{(Color online) (a)-(d): Experimentally measured strains $e_{xx}$, $e_{yy}$, $e_{xy}$ and rotation field $\omega_{xy}$ for the left dislocation in Figure \ref{fig2}(a) that has Burgers vector $(-1,\sqrt{3})/2$. (e)-(h) Strains and rotation field predicted by linear elasticity for an isolated dislocation with the same Burgers vector. (i)-(l) GPA of the off-centered linear elasticity dislocation placed on the graphene lattice. (m)-(p) Predictions of GPA for 2D dpPE. (q)-(t) Predictions of GPA for 3D dpFvKEs. (u)-(x) Predictions from hyperstress elasticity with microstructure lengths $l_1/a = 3.1$, $l_2/a = 4.1$ and $\rho_0 = -0.5$. (y) Lattice visualization of the dislocation under study. The red box marks the region of interest, of size $2a$.}
\label{fig3}
\end{figure}

\subsubsection{Comparison between experiments and theory}

Figure \ref{fig3} compares the strain and rotation fields of a dislocation with Burgers vector $(-1,\sqrt{3})/2$ (in units of the lattice spacing) calculated by fitting different theories to experimental observations of the left dislocation in Figure \ref{fig2}(a). This dislocation is well separated from other defects present in that image. The strain and rotation fields measured in experiments using GPA, panels (a)-(d), are quite different from those calculated in linear elasticity, panels (e)-(h). Panels (i)-(l) of Figure \ref{fig3} show that placing atoms on a hexagonal lattice according to the displacement vector of a dislocation given by linear elasticity\cite{footnote} produces GPA images that are close to those observed in experiments. Thus differentiating the linear elasticity displacement vector produces strains and rotations that are quite different from (and worse than!) the corresponding fields given by GPA based on the same displacement vector. 

\begin{figure}
\includegraphics[width=15cm]{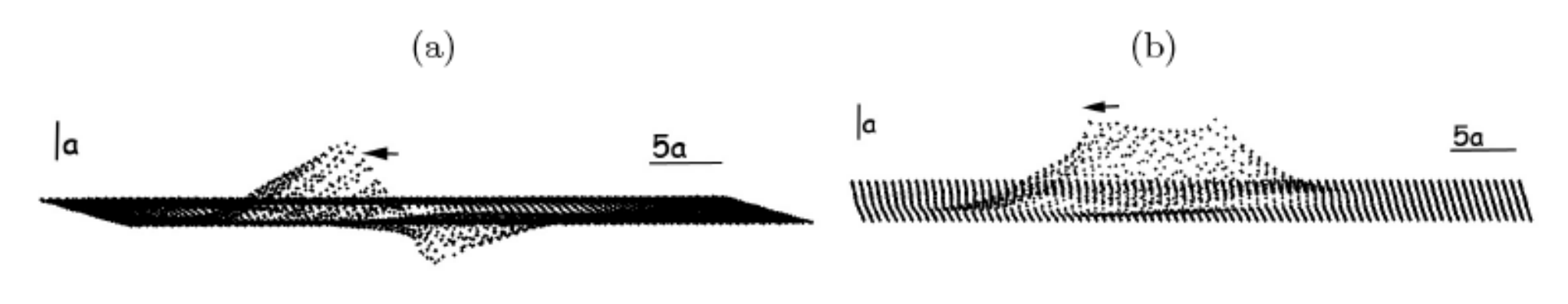}
\caption{Stationary solutions of the dpFvKEs for the dislocation dipole of figure \ref{fig2}(a). (a) Antisymmetric configuration. The left (right) dislocation in that figure sits on the peak (trough) of the rotated view in this figure. GPA of the left dislocation in (a) yields the strain and rotation fields of Figures \ref{fig3}(q)-(t). (b) Symmetric configuration. The unit of length is $a$. } 
\label{fig4}
\end{figure}

The off-centered linear elasticity dislocations relax to stationary lattice configurations when using 2D discrete periodized planar elasticity (dpPE); cf.ÊSection~\ref{sec:3}. The corresponding strain and rotation fields improve the agreement with experiments, as shown in Figures \ref{fig3}(m)-(p). Nonlinear effects for stationary dislocations in 2D dpPE are restricted to a quite small region of width smaller than $0.1a$ about the dislocation point, whereas lattice effects are negligible beyond a few lattice spacings away from it.\cite{zha06} Then two dislocations as separated as those in Figure \ref{fig2}(a) are close to the superposition of two single ones. Thus a single dislocation is a reasonable model for the left dislocation if we use 2D theories.\cite{cb05,car08} However F\"oppl-Von K\'arm\'an equations (FvKEs) and discrete periodized F\"oppl-Von K\'arm\'an equations (dpFvKEs) are uniformly nonlinear\cite{ll7,bon12,bonJSTAT12} and two dislocations placed on the graphene sheet interact and cause out-of-plane ripples and corrugations about the dislocations. Different initial dipole configurations like that in Figure \ref{fig2}(a) may evolve to four stable stationary configurations. Two of them are displayed in Figure \ref{fig4} and the other two follow from these by inverting the sign of the out-of-plane displacement. Calculating the GPA of these four configurations, we conclude that the antisymmetric configuration of Figure \ref{fig4}(a) provides the best fit to the experimental GPA of the dipole. There are two caveats to this conclusion. Firstly the calculated GPA does not include the defect region near the dipole in Figure \ref{fig2}(b), and secondly, the differences between the different calculated GPA configurations are not large. The strains and rotation of the left dislocation in Figure \ref{fig4}(a) are displayed in Figures \ref{fig3}(q)-(t). Quantitative agreement with   the experimental strains and rotation of Figures \ref{fig3}(a)-(d) is better illustrated by Figure \ref{fig6} that displays the difference between experimental measurements and theoretical predictions. Note that the antisymmetric buckling configuration of Figure \ref{fig4}(a) is similar to that in Figure 3(d) of Lehtinen et al\cite{leh13} (for well separated dislocations) although these authors considered dislocation loops (extra lines of atoms in the dislocations pointing to each other, not away from each other as in our dislocation dipole)\cite{cb05}. The shape of the symmetric configuration of Figure \ref{fig4}(b) depends on the value of $\kappa$, fitted here to $1$ eV. The  elevated intermediate region joining both dislocations almost disappears for larger $\kappa$. As $\kappa$ decreases, the curvature of the connecting region increases.

Discrete periodized planar elasticity and discrete periodized F\"oppl-Von K\'arm\'an equations produce strains and rotation that resemble experimental observations (errors are about 10\% but see later for details); cf. Figures \ref{fig3}(m)-(p) and (q)-(t), and Section \ref{sec:3}. Although atom displacements from the horizontal plane may be relatively large near defects (and may be affected by the distance to nearby defects\cite{leh13}), their projections on the plane as seen by GPA are small and therefore dpPE gives strains and rotation close to observations. Gradient elasticity such as 2D Mindlin's hyperstress theory\cite{min64} (in which stress is a function of strains and gradients of strains and rotations) gives useful explicit formulas for strain and rotation fields (see Section \ref{sec:4}) that, once calibrated, reasonably fit observations, as illustrated by panels (u)-(x) of Figure \ref{fig3}. 2D theories produce symmetric strains and rotation (via explicit expressions in the case of hyperstress theory) whereas fields from experiments display asymmetries that may indicate nonplanar effects captured by discrete periodized F\"oppl-Von K\'arm\'an equations. 

\begin{figure}
\includegraphics[width=15cm]{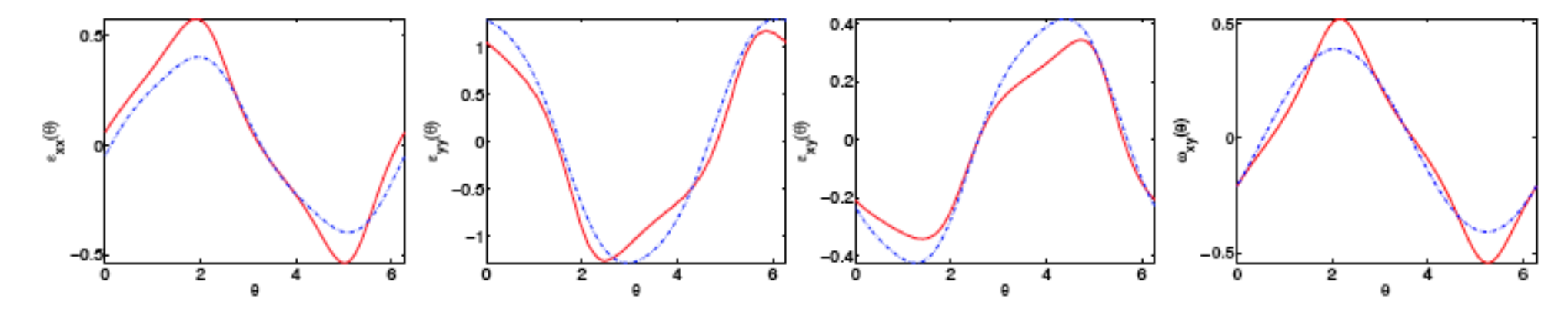}
\caption{(Color online) Theoretical strain and rotation fields (hyperstress elasticity: blue dashed lines) on circumferences of radius $R=0.05$ nm ($0.2\, a$) compared to the experimental fields (red solid line). Parameter values as in Figure \ref{fig3} and angles measured in radians. } 
\label{fig5}
\end{figure}

Figure \ref{fig5} shows that hyperstress elasticity predicts strain and rotation data that agree with experimentally measured ones on a circle about the dislocation point. This modification of conventional elasticity thus fits experimental measurements and yields values of additional elastic moduli via two microscopic lengths $l_1$ and $l_2$. At distances much larger than these lengths, hyperstress theory becomes conventional isotropic elasticity. The length scale $l_1$ characterizes the dilatation at the dislocation core, $l_2$ characterizes the rotation. In graphene they are 3.1 and 4.1 times the lattice spacings, respectively, which explains why the measured elastic fields at dislocation cores are so different from conventional elasticity predictions. We have verified these values of the length scales using circumferences of different radii\cite{radii} and also other dislocations present in the experimental data; see Section \ref{sec:4}. We have also checked that the strains and rotation of dislocation solutions of 2D discrete periodized planar elasticity, 3D discrete periodized F\"oppl-Von K\'arm\'an equations and the hyperstress expressions qualitatively agree with each other for different Burgers vectors.

Strictly speaking, our hyperstress potential energy (cf.\ Section \ref{sec:4}) is the small wavenumber and frequency limit of more complex hyperstress theories\cite{min64}. The latter consider that the solid has a microstructure that undergoes rotations and deformations, affects and is affected by the macroscopic strain. The dispersion relation has optical and acoustic branches and, in the latter's zero-frequency limit, the microscopic displacement gradients are enslaved by the macroscopic displacement. Dislocation motion in graphene is so slow that the dislocation field is stationary, hence well described by the zero-frequency equations used here.

\begin{figure}
\includegraphics[width=15cm]{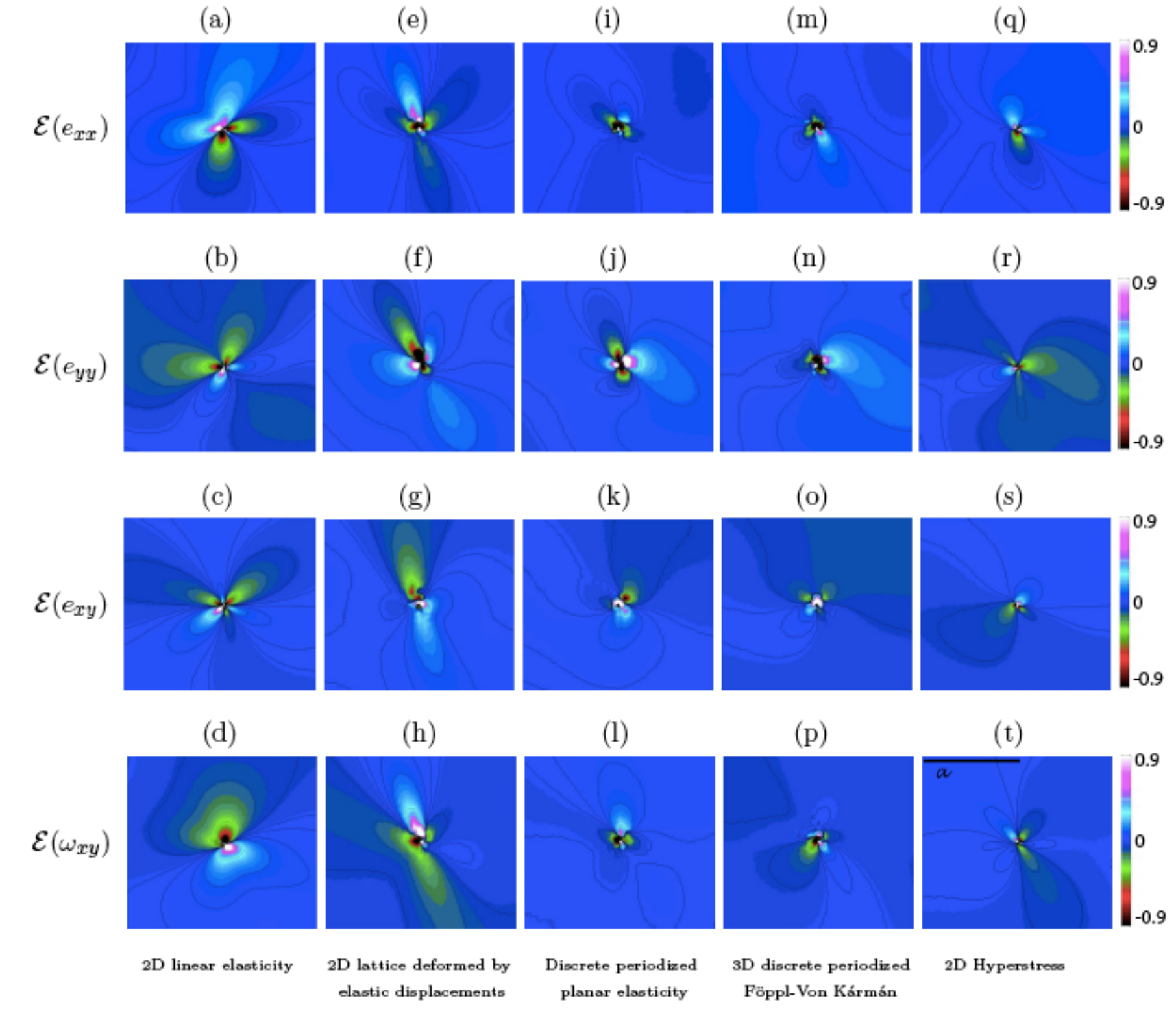} 
\caption{(Color online) Errors comparing the experimentally measured strains $e_{xx}$, $e_{yy}$, $e_{xy}$ and rotation field for the same dislocation as in Figure \ref{fig3} to: (a)-(d) Planar linear elasticity. (e)-(h) GPA from an off-centered 2D linear elasticity dislocation placed on the graphene lattice. (i)-(l) GPA from 2D dpPE after short time relaxation. (m)-(p) GPA from 3D dpFvKEs. (q)-(t) Hyperelasticity with the same parameters as in Figure \ref{fig3}. }  \label{fig6}
\end{figure}

Figure \ref{fig6} depicts the differences between experimentally measured and theoretically calculated strain and rotation fields. As anticipated from Figure \ref{fig3}, Panels (a)-(d) of Figure \ref{fig6} show that errors are large if the comparison is made with conventional 2D linear elasticity formulas. Figures \ref{fig3}(i)-(l) show that simply placing atoms on the hexagonal lattice at positions given by the displacement vector of a dislocation calculated from conventional elasticity formulas\cite{footnote} produces (by GPA) strains and rotation that are closer qualitatively to experimental measurements. However, this improvement is not evident from the errors shown in Figures \ref{fig6}(e)-(h). Time evolution of the atoms following 2D discrete periodized planar elasticity (dpPE) or 3D discrete periodized F\"oppl-Von K\'arm\'an equations (dpFvKEs) with an initial dipole configuration produce configurations whose respective strains and rotation depicted in Figures \ref{fig6}(i)-(l) and \ref{fig6}(m)-(p) appreciably reduce the error. dpPE equations and dpFvKEs have to be solved numerically to obtain atom positions from which GPA extracts strain and rotation fields. As explained in Section \ref{sec:5}, calculating the differences between images obtained from discrete theories, such as dpPE and dpFvKEs, and experimental images introduces additional numerical interpolation errors. The displacement vector, strains and rotation of a dislocation can be explicitly calculated in hyperstress elasticity, as shown in Section \ref{sec:4}. These explicit formulas give a fair approximation while avoiding to use numerical simulations, cf. Figures \ref{fig6}(q)-(t). The main differences come from the fact that hyperstress elastic fields have symmetric maxima and minima whereas experimentally measured fields and results from an off-centered dislocation in a linear elasticity dislocation placed on the hexagonal lattice are asymmetric. Fields from 2D dpPE become more symmetric with time whereas they remain asymmetric when 3D dpFvKEs are numerically solved. 

\begin{figure}[h!]
\includegraphics[width=12cm]{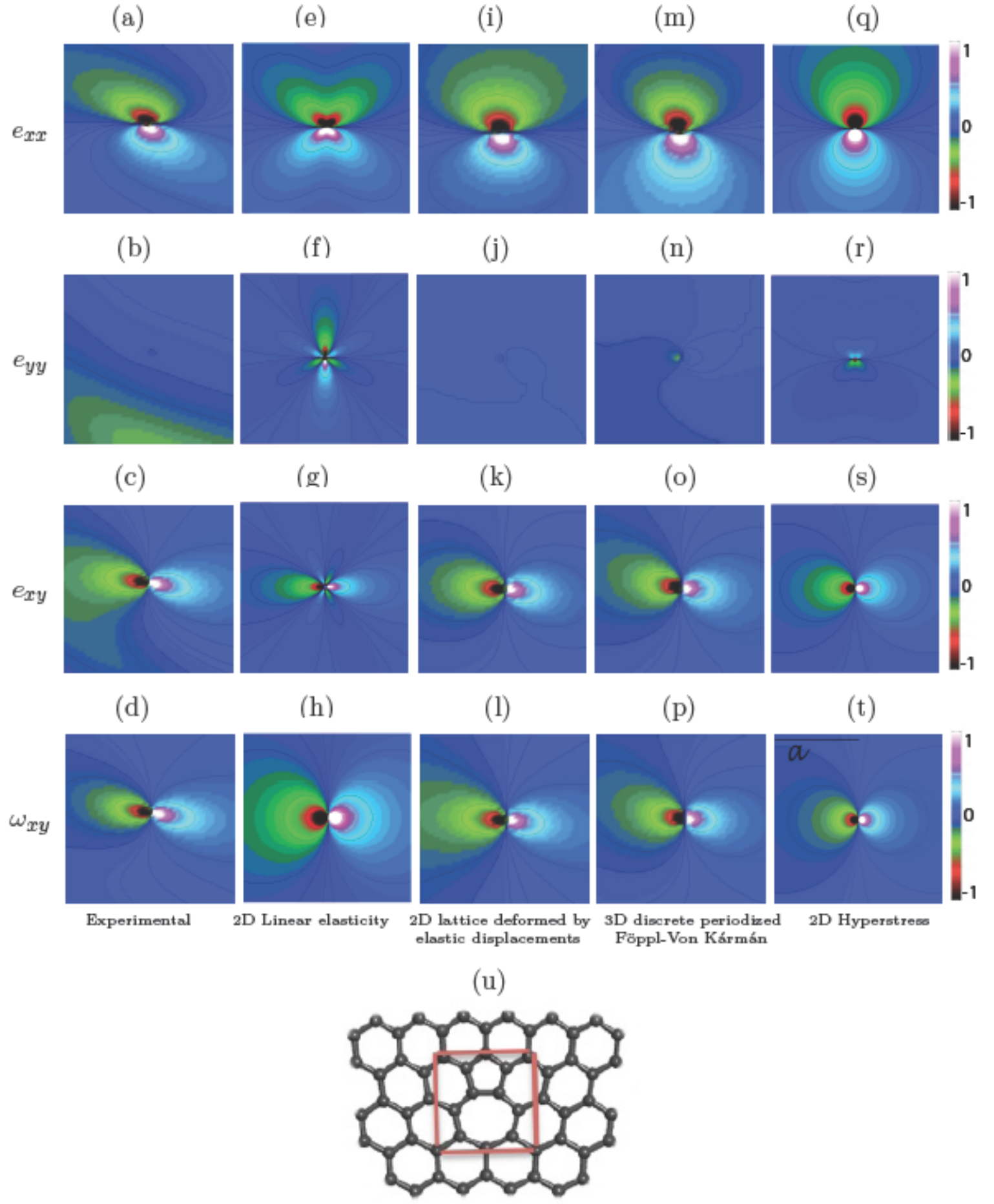} 
\caption{(Color online) (a)-(d) Experimentally measured strains $e_{xx}$, $e_{yy}$, $e_{xy}$ and rotation $\omega_{xy}$ for the top dislocation in Figure \ref{fig2}(c) that has Burgers vector $(1,0)$. (e)-(h) Strains and rotation field predicted by linear elasticity for a dislocation with the same Burgers vector. (i)-(l) GPA from an off-centered linear elasticity dislocation placed on the graphene lattice. (m)-(p) Predictions of GPA from dpFvKEs for the antisymmetric configuration of a dislocation pair. (q)-(t) Predictions from hyperstress elasticity with the same parameters as in Figure \ref{fig3}. 
(u) Lattice visualization of the dislocation. The red box marks the region of interest, of size $2a$.}
\label{fig7}
\end{figure}

\begin{figure}
\includegraphics[width=15cm]{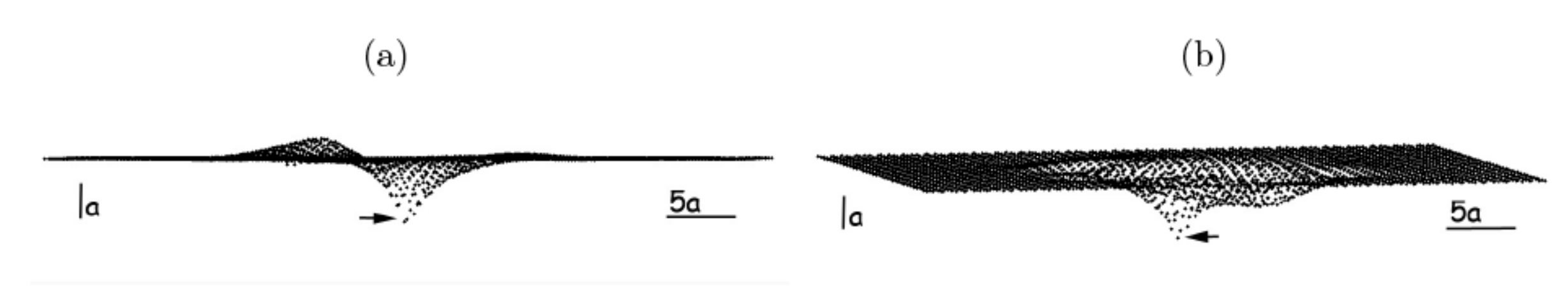}
\caption{Stationary solutions of the dpFvKEs for the dislocation pair of figure \ref{fig2}(c). (a) Antisymmetric configuration. GPA of the dislocation marked with an arrow in (a) yields the strain and rotation fields of Figures \ref{fig7}(m)-(p). (b) Symmetric configuration. The arrows in these rotated views mark the upper dislocation in Figure \ref{fig2}(c). The unit of length is $a$. }\label{fig8}
\end{figure}

\subsubsection{Data from other dislocations}

Strain and rotation fields for the upper dislocation with Burgers vector $(1,0)$ of Figure \ref{fig2}(c) are shown in Figures \ref{fig7}(a)-(d). This dislocation is separated from other defects present in the image. Compared to these fields, those predicted by 2D hyperstress elasticity shown in Figures \ref{fig7}(q)-(t) are more symmetric. Placing an off-centered linear elasticity dislocation on the graphene lattice produces tilted fields as in Figures \ref{fig7}(i)-(l), whereas stable solutions of the 
discrete periodized F\"oppl-Von K\'arm\'an equations (dpFvKEs) yield the improved fields of Figures  
\ref{fig7}(m)-(p). Calculation of errors show that the the numerical solution of the dpFvKEs gives the results that are closer to experimental values. Solving the dpFvKEs with different initial dislocation pair configurations like that in Figure \ref{fig2}(c), we find that these initial conditions may evolve to four stable stationary configurations. Two of them are displayed in Figure \ref{fig8} and the other two follow from these by inverting the sign of the out-of-plane displacement. Calculating the GPA of these four configurations, we conclude that the antisymmetric configuration of Figure \ref{fig8}(a) provides the best fit to the experimental GPA of the pair. There are two caveats to this conclusion. Firstly the calculated GPA does not include the defect region near the pair in Figure \ref{fig2}(d), and secondly, the differences between the different calculated GPA configurations are not large. The strains and rotation corresponding to the upper dislocation in Figure \ref{fig2}(c) are displayed in Figures \ref{fig7}(m)-(p). 

The out-of-plane displacements of the symmetric and antisymmetric solutions of dpFvKEs correspond to different in-plane displacements of the carbon atoms. The corresponding two dimensional lattice projections differ, and GPA may be able to distinguish them. Comparing the GPA of the 2D projections to the experimental GPA, we may establish which one is closer, thereby extracting information about out-of-plane displacements from observed in-plane displacements.

\begin{figure}
\centering
\includegraphics[width=14cm]{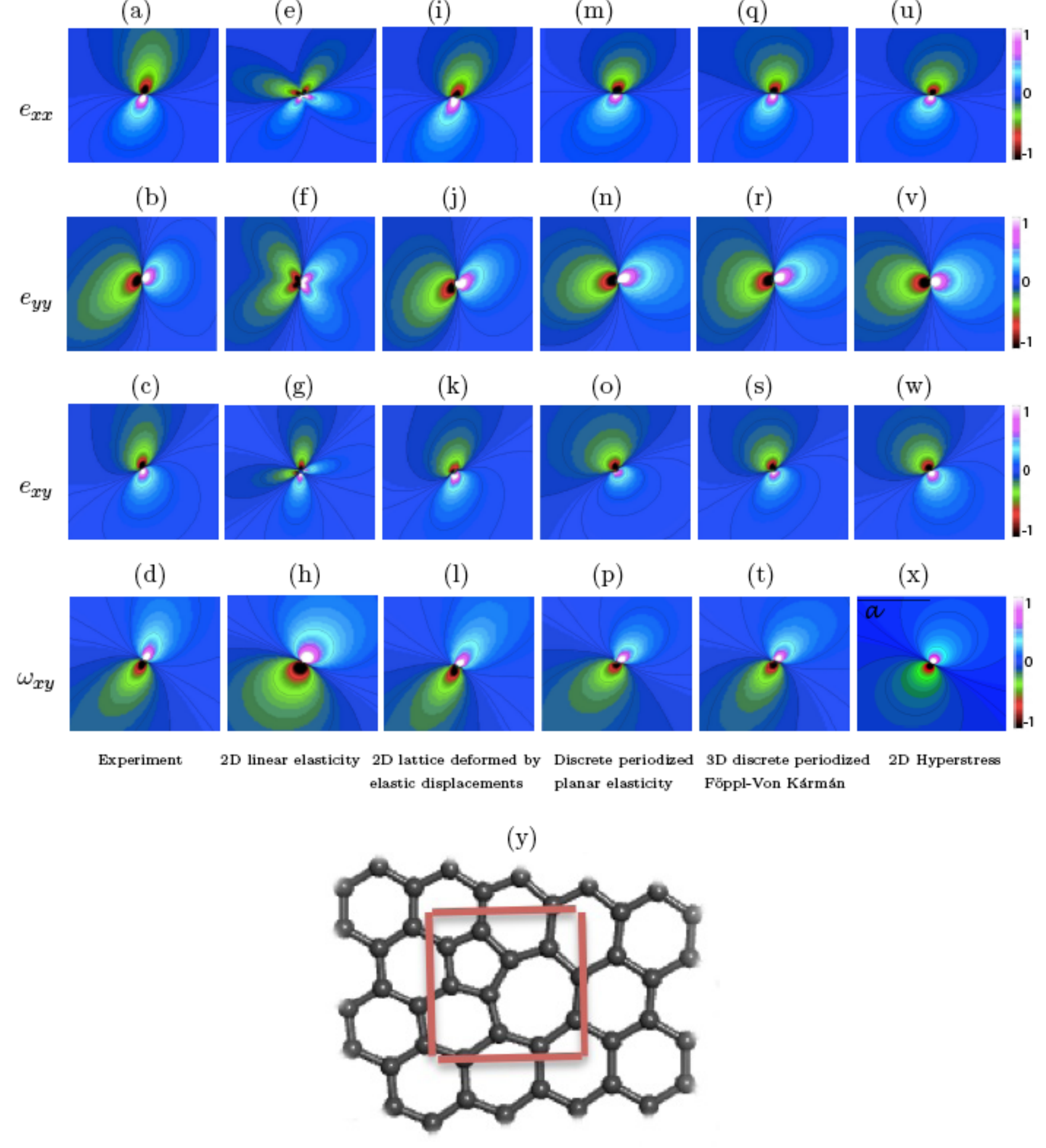}
\caption{(Color online) (a)-(d): Experimentally measured strains $e_{xx}$, $e_{yy}$, $e_{xy}$ and rotation field $\omega_{xy}$ for the dislocation in Figure \ref{fig2}(e) with Burgers vector $(1,\sqrt{3})/2$. For an isolated dislocation with the same Burgers vector, predicted strains and rotation by: (e)-(h) linear elasticity; (i)-(l) GPA of the off-centered linear elasticity dislocation placed on the graphene lattice; (m)-(p) GPA for 2D dpPE; (q)-(t) GPA for 3D dpFvKEs; (u)-(x) hyperstress elasticity (with the same parameters as in Figure \ref{fig3}).  (y) Lattice visualization of the dislocation under study. The red box marks the region of interest, of size $2a$.}
\label{fig9}
\end{figure}

Figures \ref{fig9}(a)-(d) show the experimentally measured strain and rotation fields for the isolated dislocation with Burgers vector $(1/2, \sqrt{3}/2)$ of Figure \ref{fig2}(e). These figures can be analyzed similarly to the other dislocations we have discussed. For the isolated dislocation with Burgers vector $(1/2, \sqrt{3}/2)$, the predictions of linear elasticity are depicted in Figures \ref{fig9}(e)-(h), GPA of the off-centered linear elasticity dislocation placed on the graphene lattice in Figures \ref{fig9}(i)-(l), GPA for 2D discrete periodized planar elasticity in Figures \ref{fig9}(m)-(p),  GPA for 3D discrete periodized F\"oppl-Von K\'arm\'an equations in Figures \ref{fig9}(q)-(t), and hyperstress theory in Figures \ref{fig9}(u)-(x). The differences between the different discrete predictions, the hyperstress theory and the experimental measurements are qualitatively and quantitatively small, as shown by computing the errors the same way as in Figure \ref{fig6}. Hyperstress and 3D predictions seem to provide the best overall fit.

\begin{figure}
\includegraphics[width=15cm]{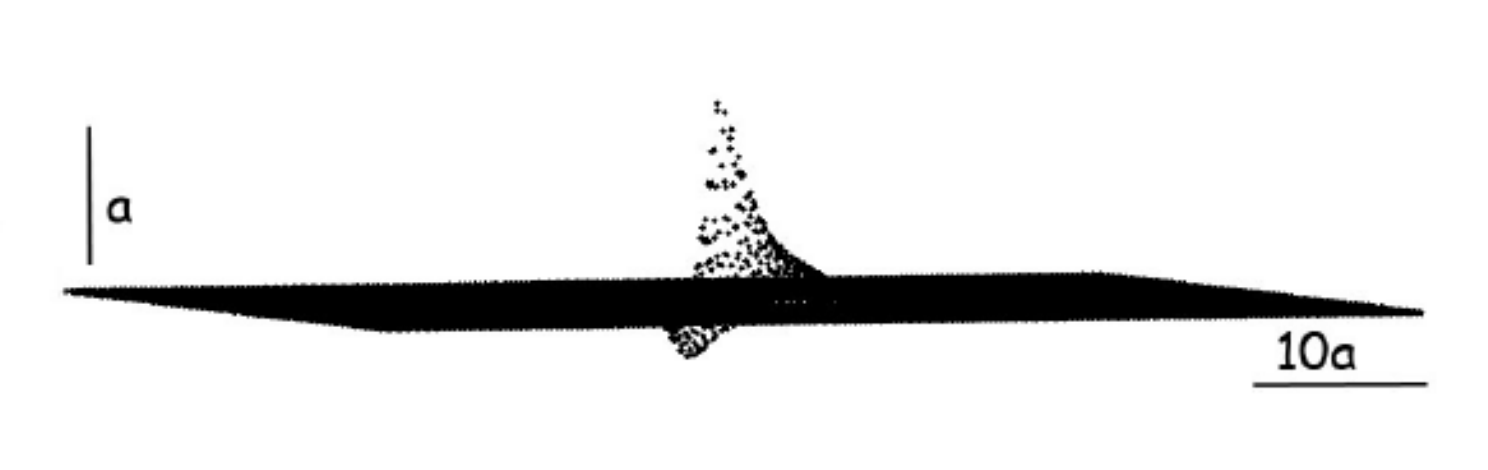}
\caption{Stationary solution of the dpFvKEs for the single dislocation of figure \ref{fig2}(e). The unit of length is $a$.}
\label{fig10}
\end{figure}

Figure \ref{fig10} represents the stationary configuration reached by a single dislocation governed by discrete periodized F\"oppl-Von K\'arm\'an equations. This configuration has been calculated under zero external stress. A second configuration obtained by changing the sign of the out-of-plane displacements is also a solution of the same equations. Both correspond to the same in-plane displacements and are therefore indistinguishable by GPA, unless some tension is present in the lattice. Under tension, a dislocation buckled upwards and a dislocation buckled downwards develop different in-plane displacements, and could be distinguished by GPA of their lattice images. 

\subsubsection{Spurious effects}
\begin{figure}
\centering
\includegraphics[width=15cm]{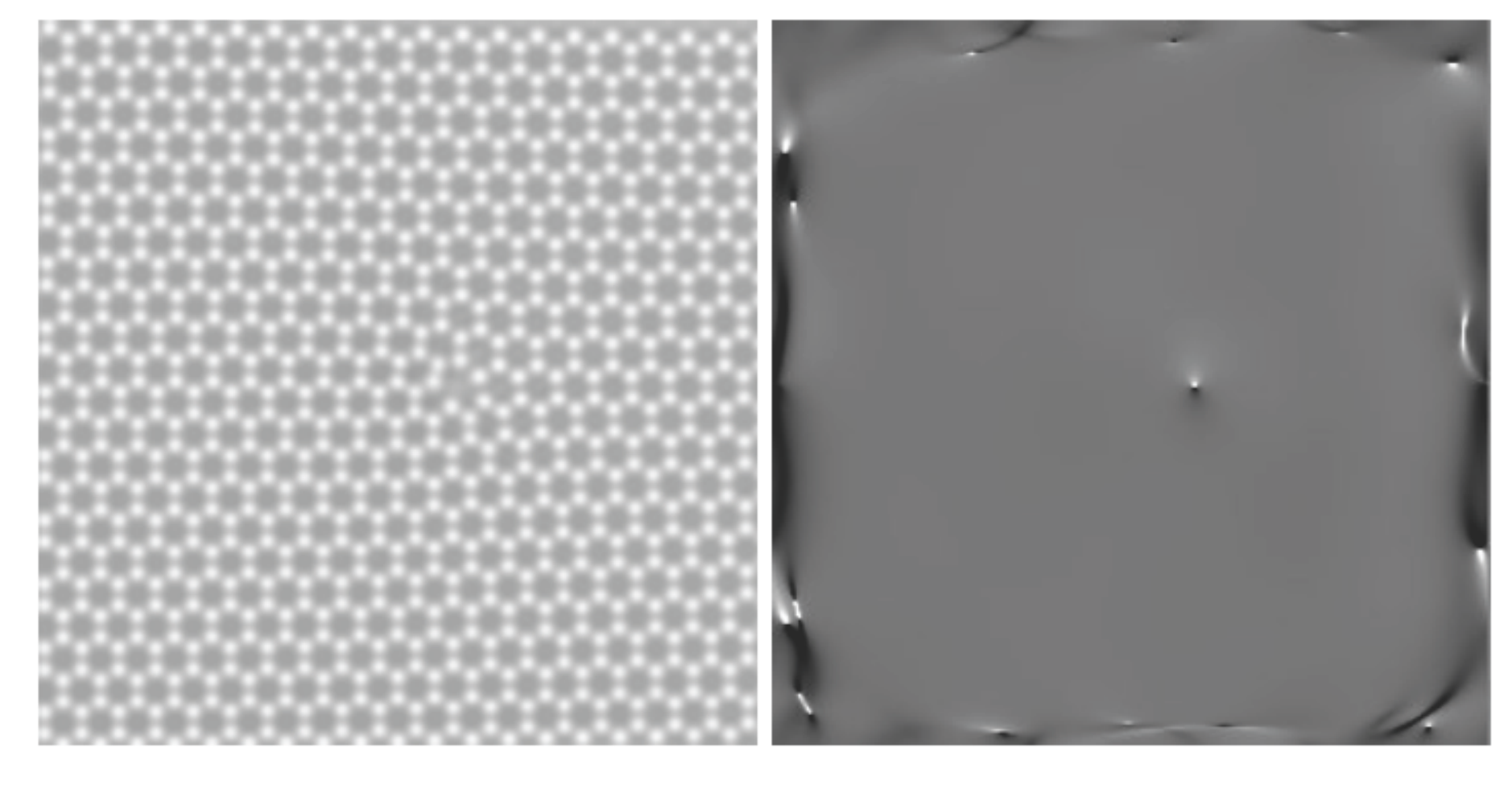}
\caption{(a) Multislice image simulation of atomic structure based on the atomic coordinates numerically generated using periodized 2D dpPE. Dislocation has Burgers vector $(-1,\sqrt{3})/2$.  (b) Strain field $e_{xx}$ corresponding to (a) generated by GPA. The only real defect is the dislocation in the middle of the figure. The spurious defects at the border of the lattice are due to the finite size of the image.}\label{fig11}
\end{figure}

GPA locates precisely dislocation points and small groupings of separated defects and gives their Burgers vectors  through their strain and rotation fields. When compared to stable solutions of the discrete periodized F\"oppl-Von K\'arm\'an equations, it helps ascertaining possible 3D configurations of the atoms at dislocation cores. However, we should be careful when using GPA as its results are quite sensitive to disturbances. In addition to the uncertainty about the presence of far away defects, we should be aware of possible spurious effects in experimental and numerical data. In fact, selecting a finite portion of the lattice introduces spurious defects in the strain and rotation fields at the boundary of the image, as it is clearly appreciated in Figure \ref{fig2}(d). This is made clear by processing lattices with a single numerically generated dislocation. Figure \ref{fig11}(a) shows the coordinates of atoms in the field of a dislocation with Burgers vector $(-1,\sqrt{3}/2)$ that have been numerically generated by using periodized 2D discrete periodized planar elasticity. This dislocation is simply a version of the left one in Figure \ref{fig2}(a) but we have eliminated all far away defects that are present there. Thus there is only one dislocation in the lattice of Figure \ref{fig11}(a) but spurious defects appear in the GPA-generated strain field of Figure \ref{fig11}(b). 

\section{GPA Technique to generate displacement, strain and rotation fields from a lattice
image}
\label{sec:GPA}

Geometrical phase analysis (GPA) describes how the spatial frequency components (interference fringes) of  the HRTEM image vary across the field of view. The squared modulus of the Fourier transform of a HRTEM lattice image exhibits bright spots corresponding to strong Bragg-reflections. These strong frequency components provide the location of vectors of the reciprocal lattice and are related to the two-dimensional unit cell describing the projected crystalline structure. While a perfect crystal lattice gives rise to sharply peaked frequency components, variations in lattice spacings produce diffuse intensities in the Fourier transform centered around the frequencies corresponding to the mean lattice spacings. The phases of frequency components vary across the image and provide strain, rotation and displacement fields there.\cite{hyt98,hrem} GPA of a HRTEM image goes as follows:\cite{hyt98}
\begin{itemize}
\item Choose two bright spots in the power spectrum corresponding to two non-collinear reciprocal lattice vectors $\mathbf{g}_1$ and $\mathbf{g}_2$. In practice, the lattice fringes giving the
best signal-to-noise are chosen.
\item Select a region of interest that includes the spots and their surroundings with the reciprocal vectors $\mathbf{g}_1$ and $\mathbf{g}_2$. By convolving each region with a Gaussian mask, we isolate the phase of its image Fourier component. We want to calculate the displacement vectors in strained regions with respect to an unstrained reference area. Therefore we also choose such a reference area and recalculate the image phases of the strained regions referred to those of the unstrained area. The relative image phases are
\begin{equation}
P_{\mathbf{g}_j}=-2\pi\mathbf{g}_j\cdot\mathbf{u}(\mathbf{x}),\quad j=1,2,\label{gpa1}
\end{equation}
where $\mathbf{u}(\mathbf{x})=(u(\mathbf{x}),v(\mathbf{x}))$ is the displacement vector at the position $\mathbf{x}=(x,y)$ and the Fourier component is $2A_{\mathbf{g}_j}\cos[2\pi\mathbf{g}_j\cdot\mathbf{x}+P_{\mathbf{g}_j}]$.
\item The displacement field can be determined from the image phases as
\begin{equation}
\mathbf{u}(\mathbf{x})= -\frac{1}{2\pi}[P_{\mathbf{g}_1}(\mathbf{x})\mathbf{a}_1+P_{\mathbf{g}_2}(\mathbf{x})\mathbf{a}_2],\label{gpa2}
\end{equation}
where $\mathbf{a}_1$ and $\mathbf{a}_2$ are the the real-space basis vectors corresponding
to the reciprocal lattice defined by $\mathbf{g}_1$ and $\mathbf{g}_2$ (i.e., $\mathbf{a}_i\cdot\mathbf{g}_j =\delta_{ij}$; defining matrices $A$ and $G$ as having column vectors $\mathbf{a}_j$ and $\mathbf{g}_j$, respectively, we have $G^T=A^{-1}$).
\item Strains $e_{ij}$ and rotations $\omega_{ij}$ are determined by differentiating $\mathbf{u}(\mathbf{x})$ as in linear elasticity. The distortion tensor $\beta_{ij}$ can be directly obtained from numerically calculated derivatives of the image phases $P_{\mathbf{g}_j}$ by using\cite{hyt98}
\begin{eqnarray}
&& (\beta)=\left(\begin{array}{cc}
\frac{\partial u}{\partial x}&\frac{\partial u}{\partial y}\\
\frac{\partial v}{\partial x}&\frac{\partial v}{\partial y}\\
\end{array}\right)= -\frac{1}{2\pi}\left(\begin{array}{cc}
a_{1 x}& a_{2x}\\
a_{1 y}& a_{2 y}\\
\end{array}\right) \left(\begin{array}{cc}
\frac{\partial P_{\mathbf{g}_1}}{\partial x}&\frac{\partial P_{\mathbf{g}_1}}{\partial y}\\
\frac{\partial P_{\mathbf{g}_2}}{\partial x}&\frac{\partial P_{\mathbf{g}_2}}{\partial y}\\
\end{array}\right)\!, \label{gpa3}\\
&&e_{ij}= \frac{1}{2}(\beta_{ij}+\beta_{ji}),\quad \omega_{ij}=\frac{1}{2}(\beta_{ij}-\beta_{ji}).\label{gpa4}
\end{eqnarray}
The more developed calculations for large deformations are not often necessary (see Appendix E in Ref.~\onlinecite{hyt98}) though they are implemented in the software.\cite{hrem}
\end{itemize}
The image phase has apparent discontinuities where it changes abruptly from $-\pi$ to $\pi$. Thus the gradients in (\ref{gpa4}) are really calculated by creating the complex image, $e^{i P_{\mathbf{g}}(\mathbf{x})}$, and then taking the gradient $\nabla P_{\mathbf{g}}=$ Im$[e^{-i P_{\mathbf{g}}(\mathbf{x})}\nabla e^{i P_{\mathbf{g}}(\mathbf{x})}]$.\cite{hyt98} In this way, strains and rotations are free from apparent phase discontinuities and are easier to interpret than the displacement vector (\ref{gpa2}). Note that displacements, strains and rotations are obtained with respect to a reference configuration in the image itself whereas they are defined with reference to a strain-free configuration in elasticity. More information about GPA and existing software for HRTEM, users' manuals, scripting capabilities and references can be found at the web site.\cite{hrem} 

\section{Discrete periodized planar elasticity and 3D F\"oppl-von K\'arm\'an equations}
\label{sec:3} 
The free energy of a membrane subject to out-of-plane bending is 
\begin{eqnarray}
&&F_g= \frac{1}{2}\int [\kappa(\nabla^2w)^2 + (\lambda u_{ii}^2 + 2\mu u_{ik}^2)]\, dx\, dy, \label{eq1}\\
&& u_{ik}= \frac{1}{2}(\partial_{x_k}u_{i}+\partial_{x_i} u_k+\partial_{x_i}w\partial_{x_k}w),\, i,k=1,2.  \label{eq2}
\end{eqnarray}
Here $(u_1,u_2)=(u(x,y),v(x,y))$, $w(x,y)$, $\lambda$, $\mu$ and $\kappa$ are the in-plane displacement vector, the membrane vertical deflection, the two 2D Lam\'e moduli  and the bending stiffness, respectively. From (\ref{eq1}), we obtain the dynamic F\"oppl-Von K\'arm\'an equations (FvKEs)
 \begin{eqnarray}
 \rho_2 \partial_t^2 u+\gamma\partial_t u\! &=&\! \lambda\,\partial_x\left( \partial_x u + \partial_y v+\frac{|\nabla w|^2}{2}\right) + \mu\,\partial_x [2 \partial_x u + (\partial_xw)^2]\nonumber\\
 &+&\! \mu\,\partial_y\left( \partial_y u + \partial_x v+ \partial_xw\partial_y w \right), \label{eq3}\\
 \rho_2 \partial_t^2 v+\gamma\partial_t v\! &=&\! \lambda\,\partial_y\left( \partial_x u + \partial_y v+\frac{|\nabla w|^2}{2}\right) + \mu\,\partial_y[2 \partial_y v + (\partial_yw)^2]\nonumber\\
 &+&\! \mu\,\partial_x\left( \partial_y u + \partial_x v+ \partial_xw\partial_y w \right),  \label{eq4}
  \end{eqnarray}
\begin{eqnarray}
&&  \rho_2 \partial_t^2w+\gamma\,\partial_t w =\lambda\,\partial_x\left[\left( \partial_x u + \partial_y v+\frac{|\nabla w|^2}{2}\right)\partial_xw\right]\nonumber\\
&&+ \lambda\,\partial_y\left[\left( \partial_x u + \partial_y v+\frac{|\nabla w|^2}{2}\right)\partial_yw\right]\nonumber\\
&&+ \mu\,\partial_x\left\{2 \partial_x u\partial_xw +(\partial_y u + \partial_x v)\partial_yw + |\nabla w|^2\partial_xw \right\} \nonumber\\
&&+ \mu\,\partial_y\left\{( \partial_y u + \partial_x v)\partial_xw+ 2\partial_y v\partial_y w+ |\nabla w|^2\partial_yw\right\}-\kappa\, (\nabla^2)^2w, 
\label{eq5}
\end{eqnarray}
to which we have added friction terms to allow given initial configurations to relax to stable stationary ones. Here $\rho_2$ is the 2D mass density (mass per unit area). For $w=0$, these equations become those of 2D planar elasticity. 

Discrete periodized planar elasticity (dpPE),\cite{car08,CBJV08,bon11} is as follows. We discretize (\ref{eq3}) and (\ref{eq4}) with $w=0$ on the honeycomb lattice, periodize finite differences along primitive directions and find the stationary configurations. To this end, we use a numerical relaxation method consisting of setting $\rho_2=0$ in the equations and solving the resulting system with {\em initial and boundary conditions} on a finite lattice given by the exact 2D dislocation field\cite{ll7} 
\begin{eqnarray}
\mathbf{u}\equiv (u,v)=\mathbf{b}\frac{1}{2\pi}\tan^{-1}\frac{y}{x}+(b_2,-b_1)\,\frac{1-\nu}{4\pi}\ln r+\frac{1+\nu}{4\pi r^2}(b_1y-b_2x)\mathbf{x},\label{eq6}
\end{eqnarray}
for a single dislocation or by a superposition of such fields for dislocation groupings. Here $\mathbf{x}=(x,y)$ and $r=\sqrt{x^2+y^2}$. We use this dislocation field centered at dislocation points that do not coincide with any lattice point. Our choice of boundary conditions is consistent with the result that the elastic fields of dislocations in graphene decay rapidly to those given by continuum elasticity.\cite{zha06} To reduce the effects of boundary conditions, we usually work with lattices larger than $100\times 100$ lattice spacings.

In discrete periodized F\"oppl-Von K\'arm\'an equations (dpFvKEs),\cite{bon12,bonJSTAT12} we discretize the equations of motion (\ref{eq3})-(\ref{eq5}) on the honeycomb lattice, periodize finite differences along the lattice primitive directions and find a stationary configuration as explained above. We add a nonzero initial condition for $w(x,y)$ so as to obtain nonplanar configurations of dpFvKEs. 

Stationary solutions provide atom positions of a stationary dislocation field. We process these atomic coordinates using GPA, thereby obtaining the strain and rotation fields depicted in Figure \ref{fig3}. We can obtain the strains and rotation of dislocation pairs or groupings by superposing the displacement vectors (\ref{eq6}) of different single dislocations centered at appropriate points.\cite{car08,CBJV08,bon11} 

Note that the FvKEs are nonlinear due to the quadratic terms $\frac{1}{2}\partial_{x_i}w\partial_{x_k}w$ in (\ref{eq2}), whereas their absence in (\ref{eq3}) and (\ref{eq4}) when $w=0$ renders these equations linear. As a consequence, superposing displacement vector fields of different dislocations produces the field of a dislocation grouping in 2D planar elasticity, but a superposition of dislocations can be quite different from a dislocation grouping in FvKEs. This is also the case with dpPE and dpFvKEs. Periodizing discrete elasticity introduces nonlinearities and nonconvexities that are localized in a small neighborhood of the dislocation point. However, superposing dislocations still produces dislocation groupings that are very similar to the initial superposition. This is definitely not true for the dpFvKEs: due to the quadratic terms in (\ref{eq2}), dynamics converts the initial configuration in a nonplanar stable one whose $w$-displacement may be quite different. We have used different initial configurations with one or more dislocations (e.g., two forming a dipole) of different Burgers vectors on a graphene membrane that may have different types of corrugations. For a single dislocation, a stable configuration consists of a large peak and a smaller valley nearby. The transformation $w\to -w$ leaves (\ref{eq5}) invariant. Thus a symmetric configuration exchanging the roles of peak and valley (deep trough below the horizontal and a smaller hill above the horizontal) is also stable. In the case of a dipole, we can adopt two symmetric initial configurations with both dislocations having a peak or a deep trough and two asymmetric configurations (left dislocation on the peak, right one in the trough, or the opposite configuration exchanging the roles of left and right dislocations). By numerically solving the dpFvKEs, we observe how the two dislocations interact, changing the membrane curvature, until a stable buckled configuration emerges (Figure \ref{fig4}). Under zero applied tension, there are two such configurations (plus the mirror images of each one) and different initial conditions lead to one or the other. The asymmetric configuration selected for Figure \ref{fig4}(a) yields a better agreement with the strains and rotation observed in experiments (Figure \ref{fig3}). Including tension as a source term in (\ref{eq5}) breaks down the mirror reflection symmetry and allows distinguishing buckling upwards and downwards from the different in-plane displacements. Our dynamic simulations of dpFvKEs are an alternative to density functional theory calculations\cite{lee14}, or to the energy calculations of Ref.~\onlinecite{leh13} (corresponding to dislocation loops, not to the dipoles of the present work).

In our simulations, we have used a bending stiffness $\kappa= 1$ eV and 300 K Lam\'e moduli\cite{zak09} $\mu= 9.95$ eV/\AA$^2$, $\lambda+2\mu=22.47$ eV/\AA$^2$, that are compatible with experimentally measured values\cite{lee08}. As the bending stiffness may get renormalized by  quantum effects,\cite{san11} we have repeated our dynamic simulations using different values of $\kappa$ and obtained similar configurations. 

\section{Hyperstress theory and elastic fields} \label{sec:4} 
2D discrete periodized planar elasticity and 3D discrete periodized F\"oppl-Von K\'arm\'an equations are useful to interpret the results of experiments but require numerical simulations to produce the elastic fields. Since having explicit formulas avoids numerical simulations, we are interested in correcting linear elasticity to include second-order effects that may be due to the lattice in a continuum theory. Mindlin proposed a more general strain energy in linear elasticity theory dependent on strains and strain gradients \cite{min64}. The rationale of this theory is to distinguish between a microscopic displacement of the atoms within each crystal cell and a macroscopic displacement. In the limit of long wavelength and small frequency and for an isotropic crystal, the 2D strain energy density of this theory is (see Eq.\ (12.4) of Ref.~\onlinecite{min64})
\begin{equation}
W=\frac{1}{2}\int\!\left[\lambda \beta_{ii}^2 + 2\mu \beta_{(ij)}^2+ 4\eta\kappa_{i}^2+ 3a_1 \kappa_{jji}^2+2a_2\kappa_{ijk}^2+2f\epsilon_{ij}\kappa_i\kappa_{jkk}\right]\! dx\, dy. \label{eq7}
\end{equation}
Here $\beta_{ij}=\partial_j u_i=u_{i,j}$, where $u_1=u$, $u_2=v$ are the components of the displacement vector $\mathbf{u}$. Sum over repeated indices is implied. $\beta_{(ik)}\equiv (\beta_{ik}+\beta_{ki})/2$ is the elastic strain, $\beta_{[21]}\equiv (\beta_{21}-\beta_{12})/2$ is the rotation, $\kappa_i=\beta_{[21],i}$ is the curvature vector (gradient of the scalar rotation $\beta_{[21]}$), $\epsilon_{ij}$ is the completely antisymmetric unit tensor with $\epsilon_{12}=1$, and $\kappa_{ijk}=\beta_{(ij,k)}$ is the completely symmetrized tensor of the second order derivatives of $\mathbf{u}$. Stability requires $f>0$ and that the combinations
\begin{equation}
l_1^2=\frac{3a_1+2a_2}{\lambda+2\mu}, \quad l_2^2=\frac{3\eta+f+a_1+2a_2}{3\mu}, \label{eq8}
\end{equation}
with dimensions of length square be also positive\cite{min64}. Setting the variation of $W$ with respect to the displacement vector equal to zero produces the balance of momentum equation
\begin{eqnarray}
\lambda u_{j,ji}+\mu (u_{i,jj} +u_{j,ij})+\frac{4\eta+f}{2}\epsilon_{ij}\!\left(\kappa_{l,lj}+\epsilon_{lm}\kappa_{mnn,lj}\!\right)\!-a_1(\kappa_{llj,ji}+\kappa_{llj,ij}+\kappa_{lli,jj})\nonumber\\
-2a_2\kappa_{ijk,jk}-\frac{f}{3}(\epsilon_{li}\kappa_{l,jj}+2\epsilon_{lj}\kappa_{l,ji})=0. \label{eq9}
\end{eqnarray}
We now rewrite this equation using the identity
\begin{eqnarray}
\epsilon_{ji}\alpha_j=\epsilon_{ji}\epsilon_{kl}\beta_{jl,k}= \beta_{ji,j}-\beta_{jj,i}, \label{eq10}
\end{eqnarray}
where $\alpha_{i}$ is the dislocation density vector
\begin{eqnarray}
 \alpha_{i}=-\epsilon_{jk}\beta_{ij,k}=\beta_{i2,1}-\beta_{i1,2}. \label{eq11}
\end{eqnarray}
This equation follows from the definition of the Burgers vector \cite{bul06,hir82}, 
\begin{eqnarray}
b_i=\oint du_i=\oint dx_j u_{i,j} =\epsilon_{kj}\int \partial_k\beta_{ij} dx\,dy,\label{eq12}
\end{eqnarray}
that implies that the displacement vector $\mathbf{u}$ is multivalued. Here the line integral is calculated on a contour encircling the dislocation point and we have used the Stokes theorem. The displacement vector of a single dislocation has a jump discontinuity as it receives a finite increment whenever we go around any closed contour encircling the dislocation point. 

\begin{figure}
\includegraphics[width=15cm]{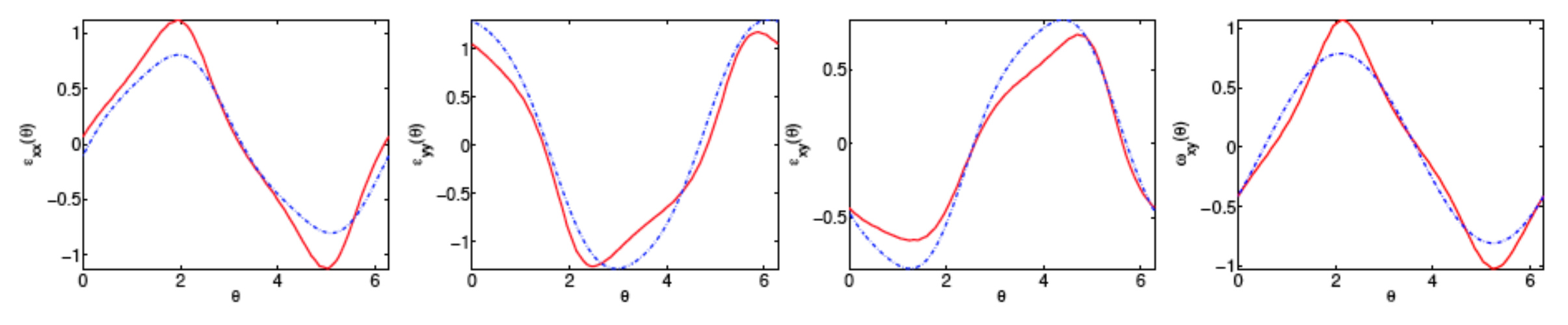}
\caption{(Color online) Theoretical strain and rotation fields (hyperstress elasticity: blue dashed lines) on circumferences of radius $R=0.025$ nm ($0.1\, a$) compared to the experimental fields (red solid line). Parameter values are $l_1/a = 3.1$, $l_2/a = 4.1$, $\rho_0 = -0.5$, and the angles are measured in radians. Data correspond to the dislocation of Figure \ref{fig3}.} 
\label{fig12}
\end{figure}

\begin{figure}
\includegraphics[width=15cm]{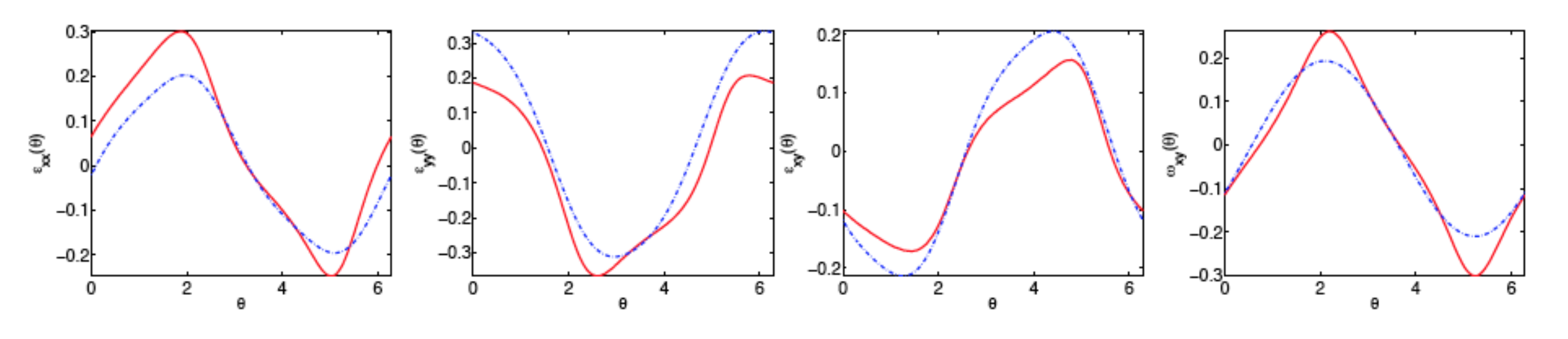}
\caption{(Color online) Same as Fig.~\ref{fig12} for $R=0.1$ nm. } 
\label{fig13}
\end{figure}

Using (\ref{eq10}), we transform (\ref{eq9}) in
\begin{eqnarray}
&&(1-l_2^2\Delta)\Delta\mathbf{u}+\left[1+\frac{\lambda}{\mu}+\left(l_2^2-\frac{2l_1^2}{1-\nu}\right)\!\Delta\!\right]\!\nabla(\nabla\cdot\mathbf{ u}) \nonumber\\
&&\quad =\!\left\{\!(b_2,-b_1)\!\left[1+\left(l_2^2-\frac{2l_1^2}{1-\nu}+\frac{2a_1}{\mu}\right)\!\Delta\!\right]\!-\frac{2a_1}{3\mu}(b_2\partial_1-b_1\partial_2)\nabla\right\}\!\delta(x)\delta(y), \label{eq13}
\end{eqnarray}
where $\Delta\mathbf{u}=\nabla^2\mathbf{u}$. Solving this equation as indicated in Appendix \ref{app} produces the displacement vector of a static dislocation with Burgers vector $\mathbf{b}=(b_1,b_2)$ centered at the origin: 
\begin{eqnarray}
\mathbf{u}&=&\mathbf{b}\frac{1}{2\pi}\tan^{-1}\frac{y}{x}+\frac{(b_2,-b_1)}{\pi}\!\left\{\frac{1-\nu}{4}\ln r+\!\left(\frac{\nu}{2}-\frac{1-\nu}{3\mu l_1^2}a_1\right)\!\left[\frac{l_1}{r} K_1\!\left(\frac{r}{l_1}\right)\!-\frac{l_1^2}{r^2}\right]\!\right.\nonumber\\
&+&\left. \!\left(1+\frac{a_1}{\mu l_2^2}-\frac{l^2_1}{(1-\nu)l_2^2}\right)\!\left[K_0\!\left(\frac{r}{l_2}\right)\!+\frac{l_2}{r}K_1\!\left(\frac{r}{l_2}\right)\!-\frac{l_2^2}{r^2}\!\right]\! \right\}\!+\mathbf{x}\frac{b_1y-b_2x}{\pi r^2}\!\left\{\frac{1+\nu}{4}\right.\nonumber\\ 
&+&\!\left(1+\frac{a_1}{\mu l_2^2}-\frac{l_1^2}{(1-\nu)l_2^2}\right)\!\left[K_2\!\left(\frac{r}{l_2} \right)\!-\frac{2l_2^2}{r^2}\right]\!+\left.\!\left(\frac{\nu}{2}-\frac{a_1(1-\nu)}{3\mu l_1^2}\right)\!\left[K_2\!\left(\frac{r}{l_1}\right)\!-\frac{2l_1^2}{r^2}\right]\!\right\}\!\!. \label{eq14}
\end{eqnarray}
Here $K_\nu(x)$ are modified Bessel functions \cite{AS} and $r=\sqrt{x^2+y^2}$. The length $l_2$ and the parameter $\rho_0=\frac{a_1}{\mu l_2^2}-\frac{l_1^2}{(1-\nu)l_2^2}$ control the rotation field whereas $l_1$ is related to the dilatation. By fitting experimental data as shown in Figure \ref{fig5}, we have estimated the values $l_1=3.1b$, $l_2=4.1 b$, $\rho_0=-0.5$ ($b=2.46$ \AA\, is the graphene lattice spacing) which yield $a_1=163.28$ eV, $a_2=398.32$ eV and $\eta + \frac{1}{3}f= 639.95$ eV. We have used the same values of the Lam\'e moduli as in Section \ref{sec:3},\cite{zak09} $\mu= 9.95$ eV/\AA$^2$, $\lambda+2\mu=22.47$ eV/\AA$^2$. Figures \ref{fig12} and \ref{fig13} show that the same values of $l_1$, $l_2$ and $\rho_0$ fit the elastic fields on circles of radii $R=0.025$ nm and 0.1 nm about the dislocation point, respectively. Experimental data for larger radii are too noisy to be trusted. We have also checked that the same values of the parameters fit experimental data for the dislocations with Burgers vectors $\pm (1,0)$, $\pm (1,\sqrt{3})/2$, $(1,-\sqrt{3})/2$, and for the same radii as in Figures \ref{fig5}, \ref{fig12} and \ref{fig13}.

As hyperstress elasticity is a linear theory, we can obtain the displacement vectors, strains and rotation of dislocation pairs or groupings by superposing those of different single dislocations (\ref{eq14}) centered at appropriate points. The observed differences with measured values may be due to nonplanar and nonlinear effects (e.g., membrane buckling) that induce additional asymmetries and rotations with respect to the predictions of our 2D theory, or noise due to the presence of other defects and limitations of the GPA software.

The strains and rotation can be obtained from (\ref{eq14}). We first calculate its gradient,
\begin{eqnarray}
\nabla\mathbf{u}\!&=&\!\mathbf{b}(-y,x)\frac{1}{2\pi r^2}+\frac{(b_2,-b_1)\mathbf{x}}{\pi r^2}\!\left\{\frac{1-\nu}{4}-\!\left(1+\frac{a_1}{\mu l_2^2}-\frac{l^2_1}{(1-\nu)l_2^2}\right)\!\left[K_2\!\left(\frac{r}{l_2}\right)\!-\frac{2l_2^2}{r^2}\right.\right.\nonumber\\
&+&\!\!\!\left.\left.\frac{r}{l_2}K_1\!\left(\frac{r}{l_2}\right)\!\right]\!- \!\left(\frac{\nu}{2}-\frac{1-\nu}{3\mu l_1^2}a_1\right)\!\left[K_2\!\left(\frac{r}{l_1} \right)\!-\frac{2l_1^2}{r^2}\right]\!\right\}\!-\frac{\mathbf{x}(b_2,-b_1)}{\pi r^2}\left\{\frac{1+\nu}{4}\!+\!\left(1\!+\!\frac{a_1}{\mu l_2^2}\right.\right.\nonumber\\
&-&\!\!\left.\left. \frac{l^2_1}{(1-\nu)l_2^2}\right)\!\left[K_2\!\left(\frac{r}{l_2}\right)\!-\frac{2l_2^2}{r^2}\!\right]\!+\!\left(\frac{\nu}{2}-\frac{1-\nu}{3\mu l_1^2}a_1\right)\!\left[K_2\!\left(\frac{r}{l_1} \right)\!-\frac{2l_1^2}{r^2}\right]\!\right\}\!+\mathbf{I}\frac{b_1y-b_2x}{\pi r^2}\nonumber\\
&\times&\!\!\left\{\frac{1+\nu}{4}+\!\left(1+\frac{a_1}{\mu l_2^2}-\frac{l_1^2}{(1-\nu)l_2^2}\right)\!\!\left[K_2\!\left(\frac{r}{l_2} \right)\!-\frac{2l_2^2}{r^2}\right]\!+\!\left(\frac{\nu}{2}-\frac{a_1(1-\nu)}{3\mu l_1^2}\right)\!\left[K_2\!\left(\frac{r}{l_1}\right)\right.\right.\nonumber\\
&-&\! \!\left.\left.\frac{2l_1^2}{r^2}\right]\!\right\}\!-\mathbf{x}\mathbf{x}\frac{b_1y-b_2x}{\pi r^4}\!\left\{\frac{1+\nu}{2}+\!\left(1+\frac{a_1}{\mu l_2^2}-\frac{l_1^2}{(1-\nu)l_2^2}\right)\!\left[\frac{r}{l_2}K_1\!\left(\frac{r}{l_2}\right)\!+4K_2\!\left(\frac{r}{l_2} \right)\!\right.\right.\nonumber\\
&-&\!\left.\frac{8l_2^2}{r^2}\right]\!+\left.\!\left(\frac{\nu}{2}-\frac{a_1(1-\nu)}{3\mu l_1^2}\right)\!\left[4K_2\!\left(\frac{r}{l_1} \right)\!+\frac{r}{l_1}K_1\!\left(\frac{r}{l_1}\right)\!-\frac{8l_1^2}{r^2}\right]\!\right\}\!, \label{eq25}
\end{eqnarray}
thereby obtaining the following strains and rotation
\begin{eqnarray}
\beta_{11}\!&=&\!\frac{b_1y}{\pi r^2}\!\left\{\frac{1+\nu}{4r^2}(y^2-x^2)-\frac{1}{2}+\!\left(1+\frac{a_1}{\mu l_2^2}-\frac{l_1^2}{(1-\nu)l_2^2}\right)\!\left[\frac{y^2-3x^2}{r^2}\!\left(K_2\!\left(\frac{r}{l_2} \right)\!-\frac{2l_2^2}{r^2}\right)\right.\right.\nonumber\\
&-&\left.\left.\!\frac{x^2}{rl_2}K_1\!\left(\frac{r}{l_2} \right)\!\right]\!+\!\left(\frac{\nu}{2}-\frac{a_1(1-\nu)}{3\mu l_1^2}\right)\!\left[\frac{y^2-3x^2}{r^2}\!\left(K_2\!\left(\frac{r}{l_1}\right)\!-\frac{2l_1^2}{r^2}\right)\!-\frac{x^2}{rl_1}K_1\!\left(\frac{r}{l_1} \right)\!\right]\!\right\}\!\nonumber\\ 
&-&\left.\!\frac{b_2x}{\pi r^2}\!\left\{\frac{\nu}{2}+\frac{1+\nu}{4r^2}(y^2-x^2)\! \right.+\!\left(1+\frac{a_1}{\mu l_2^2}-\frac{l^2_1}{(1-\nu)l_2^2}\right)\!\left[\frac{3y^2-x^2}{r^2}\!\left(K_2\!\left(\frac{r}{l_2}\right)\!-\frac{2l_2^2}{r^2}\right)\!\right.\right.\nonumber\\
&+&\!\left.\frac{y^2}{r l_2}K_1\!\left(\frac{r}{l_2}\right)\!\right]\!+\!\left(\frac{\nu}{2}-\frac{1-\nu}{3\mu l_1^2}a_1\right)\!\left[\frac{3y^2-x^2}{r^2}\!\left(K_2\!\left(\frac{r}{l_1} \right)\!-\frac{2l_1^2}{r^2}\right)\!-\left.\frac{x^2}{rl_1}\!K_1\!\left(\frac{r}{l_1} \right)\!\right]\!\right\}\!, \label{eq26}
\end{eqnarray}
\begin{eqnarray}
\beta_{22}\!&=&\!\frac{b_1y}{\pi r^2}\!\left\{\frac{\nu}{2}+\frac{1+\nu}{4r^2}(x^2-y^2)+\!\left(1+\frac{a_1}{\mu l_2^2}-\frac{l_1^2}{(1-\nu)l_2^2}\right)\!\left[\frac{3x^2-y^2}{r^2}\!\left(K_2\!\left(\frac{r}{l_2} \right)\!-\frac{2 l_2^2}{r^2}\right)\!\right.\right.\nonumber\\
&+&\!\left.\left.\frac{x^2}{rl_2}K_1\!\left(\frac{r}{l_2} \right)\!\right]\!+\!\left(\frac{\nu}{2}-\frac{a_1(1-\nu)}{3\mu l_1^2}\right)\!\left[\frac{3x^2-y^2}{r^2}\!\left(K_2\!\left(\frac{r}{l_1}\right)\!-\frac{2l_1^2}{r^2}\right)\!-\frac{y^2}{rl_1}K_1\!\left(\frac{r}{l_1} \right)\!\right]\!\right\}\nonumber\\ 
&-&\!\frac{b_2x}{\pi r^2}\!\left\{\frac{1+\nu}{4r^2}(x^2-y^2)-\frac{1}{2}+\!\left(1+\frac{a_1}{\mu l_2^2}-\frac{l^2_1}{(1-\nu)l_2^2}\right)\!\left[\frac{x^2-3y^2}{r^2}\!\left(K_2\!\left(\frac{r}{l_2}\right)\!-\frac{2l_2^2}{r^2}\right)\!\right.\right.\nonumber\\
&-&\left.\!\frac{y^2}{r l_2}K_1\!\left(\frac{r}{l_2}\right)\!\right]\!+\!\left(\frac{\nu}{2}-\frac{1-\nu}{3\mu l_1^2}a_1\right)\!\left[\frac{x^2-3y^2}{r^2}\!\left(K_2\!\left(\frac{r}{l_1} \right)\!-\frac{2l_1^2}{r^2}\right)\!-\left.\frac{y^2}{rl_1}\!K_1\!\left(\frac{r}{l_1} \right)\!\right]\!\right\}\!, \label{eq27}
\end{eqnarray}
\begin{eqnarray}
\beta_{(12)}\!&=&\!\frac{\mathbf{b}\cdot\mathbf{x}(x^2-y^2)}{\pi r^4}\!\left\{\frac{1+\nu}{4}+ \!\left(1+\frac{a_1}{\mu l_2^2}-\frac{l_1^2}{(1-\nu)l_2^2}\right)\!\left[K_2\!\left(\frac{r}{l_2} \right)\!-\frac{2 l_2^2}{r^2}\right.\right.\nonumber\\
&+&\!\left.\!\frac{r}{2l_2}K_1\!\left(\frac{r}{l_2}\right)\!\right]\!+\!\left(\frac{\nu}{2}-\frac{a_1(1-\nu)}{3\mu l_1^2}\right)\!\left.\!\left[K_2\!\left(\frac{r}{l_1}\right)\!-\frac{2l_1^2}{r^2}+\frac{r}{2l_1}K_1\!\left(\frac{r}{l_1}\right)\!\right]\!\right\}\!+ \!\frac{b_2y-b_1x}{2\pi rl_1}\nonumber\\
&\times&\!\!\left(\frac{\nu}{2}-\frac{a_1(1-\nu)}{3\mu l_1^2}\!\right)\!K_1\!\left(\frac{r}{l_1}\right)\!+\!\frac{2xy(b_2x-b_1y)}{\pi r^4}\!\left\{\!\left(1+\frac{a_1}{\mu l_2^2}-\frac{l_1^2}{(1-\nu)l_2^2}\right)\!\!\left[K_2\!\left(\frac{r}{l_2} \right)\right.\right.\nonumber\\
&-&\left.\left.\!\!\frac{2 l_2^2}{r^2}\right]\!+\!\left(\frac{\nu}{2}-\frac{a_1(1-\nu)}{3\mu l_1^2}\right)\!\!\left[K_2\!\left(\frac{r}{l_1}\right)\!-\frac{2l_1^2}{r^2}\right]\!\right\}\!, \label{eq28}\\
\beta_{[12]}&=&\frac{\mathbf{b}\cdot\mathbf{x}}{2\pi r^2}\!\left\{1-\!\left(1+\frac{a_1}{\mu l_2^2}-\frac{l_1^2}{(1-\nu)l_2^2}\right)\!\frac{r}{l_2}K_1\!\left(\frac{r}{l_2} \right)\!\right\}\!. \label{eq29}
\end{eqnarray}
Here $\beta_{11}=e_{xx}$, $\beta_{22}=e_{yy}$, $\beta_{(12)}=e_{xy}$ are the strains and $\beta_{[12]}=\omega_{xy}$ is the rotation. The dilatation is 
\begin{eqnarray}
\beta_{11}+\beta_{22}=\frac{b_2x-b_1y}{2\pi r^2}\!\left[1-\nu+\!\left(\nu-2a_1\frac{1-\nu}{3\mu l_1^2}\right)\!\frac{r}{l_1}K_1\!\left(\frac{r}{l_1} \right)\!\right]\!. \label{eq30}
\end{eqnarray}
The modified Bessel functions in these formulas decay exponentially fast at distances from the dislocation point that are much larger than $l_1$ and $l_2$. Then strains and rotation become those of conventional elasticity. 

\section{Guide for using the different theories}\label{sec:5}
Figures \ref{fig3}, \ref{fig7} and \ref{fig9} depict strain and rotation fields calculated by the theories described in Sections \ref{sec:3} and \ref{sec:4}:
\begin{itemize}
\item The strains and rotation for linear elasticity are obtained by differentiating the displacement vector (\ref{eq6}). Equivalently, removing all terms involving modified Bessel functions $K_1$, $K_2$, and the parameters $l_1$, $l_2$ from Equations (\ref{eq26})-(\ref{eq29}) and setting $\beta_{11}=e_{xx}$, $\beta_{22}=e_{yy}$, $\beta_{(12)}=e_{xy}$ and $\beta_{[12]}=\omega_{xy}$. The level curves of strains and rotation are very different from experimental ones in Figures \ref{fig3}, \ref{fig7} and \ref{fig9}.
\item The strains and rotation for hyperstress theory are given by Equations (\ref{eq26})-(\ref{eq29}) of Section \ref{sec:4} and setting $\beta_{11}=e_{xx}$, $\beta_{22}=e_{yy}$, $\beta_{(12)}=e_{xy}$ and $\beta_{[12]}=\omega_{xy}$. Note that the level curves of strains and rotation are round and positive ones are symmetric to negative ones in Figures \ref{fig3}, \ref{fig7} and \ref{fig9}.
\item Off-centered linear elasticity strains and rotation are obtained by first replacing points $(x_i,y_i)$, $i=1,\ldots, N$ of the hexagonal lattice\cite{car08} by $(x_i+U(x_i-x_0,y_i-y_0),y_i+V(x_i-x_0,y_i-y_0))$, where $(U(x,y),V(x,y))$ is the elastic displacement vector (\ref{eq6}) and $(x_0,y_0)$ is a point (different from any lattice point) selected so that the deformed lattice contains the appropriate heptagon-pentagon defect. Then atom arrangement in the deformed lattice is visualized by standard scientific software such as Matlab and GPA produces the strain and rotation fields of the resulting image. The origin $(x_0,y_0)$ is not unique and is chosen by trial and error. As a result the quality of the approximation varies as can be appreciated in Figures \ref{fig3}, \ref{fig7} and \ref{fig9}.
\item Strains and rotation for discrete periodized planar elasticity are obtained applying GPA to images of  the deformed lattice $(x_i+u(x_i,y_i),y_i+v(x_i,y_i))$. Here $(u(x_i,y_i),v(x_i,y_i))$ is the stationary solution of (\ref{eq3}) and (\ref{eq4}) with $w=0$ after they are discretized on the hexagonal lattice and finite differences are periodized as indicated in Reference \onlinecite{car08}. The stationary solutions are found by the numerical relaxation method of Section~\ref{sec:3} with initial condition given by the above mentioned off-centered configuration $(x_i+U(x_i-x_0,y_i-y_0),y_i+V(x_i-x_0,y_i-y_0))$. For large lattices, the resulting dislocation core does not depend on the choice of $(x_0,y_0)$.
\item Strains and rotation for discrete periodized F\"oppl-von K\'arm\'an equations are obtained applying GPA to images of 2D projections $(x_i+u(x_i,y_i),y_i+v(x_i,y_i))$ of the deformed sheet $(x_i+u(x_i,y_i),y_i+v(x_i,y_i), w(x_i,y_i))$. Here $(u(x_i,y_i),\, v(x_i,y_i),\, w(x_i,y_i))$ is the stationary solution of (\ref{eq3}) - (\ref{eq5}) after they are discretized on the hexagonal lattice and finite differences are periodized as indicated in Reference \onlinecite{bon12}. The stationary solutions are found by the numerical relaxation method of Section~\ref{sec:3} with initial condition given by an off-centered configuration $(x_i+U(x_i-x_0,y_i-y_0),y_i+V(x_i-x_0,y_i-y_0),W(x_i,y_i))$. Here $W\neq 0$ at the dislocation core so as to obtain a nonplanar configuration.
\end{itemize}

To plot the differences between GPA of experimental images and the results obtained from theories (called errors in Figure \ref{fig6}), we need to understand the different data we are comparing. GPA of an image produces strains and rotation on a Cartesian grid and the maximum of $e_{xx}$ gives the approximate location of the dislocation point. Linear elasticity and hyperstress theory yield analytical formulas that can be evaluated at any point of the same grid (with the same dislocation point). GPA of off-centered elasticity, discrete periodized elasticity and discrete periodized F\"oppl-von K\'arm\'an equations produce strains and rotation on a cartesian grid that, typically, has a different step size from that of the experimental image. The location of the dislocation point is given again by the maximum of $e_{xx}$. To plot the differences in Figure \ref{fig6}, we need to interpolate data from one cartesian grid in the other one and subtract. Interpolation introduces numerical errors in the core region that are absent when comparing with analytic formulas (linear elasticity and hyperstress).

Discrete theories (off-centered elasticity, discrete periodized planar elasticity and discrete periodized F\"oppl-von K\'arm\'an equations) produce images of dislocation cores that are similar to the sketches in Figures \ref{fig3}(y), \ref{fig7}(u) and \ref{fig9}(y) except that the distances between atoms at the core are slightly different. An arbitrary choice of $(x_0,y_0)$ in off-centered elasticity results in some distances between atoms being noticeably longer or shorter. Discrete periodized planar elasticity homogenizes  distances between atoms. Discrete periodized F\"oppl-von K\'arm\'an equations modulates the 2D projections of atom distances by coupling to the out-of-plane displacements. 

To distinguish in-plane displacements from out-of-plane ones, we have to combine theoretical results and experiments. Our strategy consists of (i) generating numerically a collection of three
dimensional defects with different possible out-of-plane and in-plane configurations that solve discrete periodized F\"oppl-von K\'arm\'an plate equations, (ii) generating their strain and rotation fields with GPA, and (iii) comparing those fields to the GPA fields of the experimental
images. Among different defect configurations provided by theory, we select those providing the best agreement with experiments.

\section{Conclusions}\label{sec:6}

In summary, we have measured strain and rotation fields near isolated dislocations and dislocation pairs by using GPA of atomic resolution images obtained by an aberration-corrected TEM operating at 80 kV. We have compared them to numerical solutions of stationary discrete periodized planar elasticity (dpPE) and discrete periodized F\"oppl-Von K\'arm\'an equations (dpFvKEs) treated by GPA, to planar isotropic hyperstress and to conventional elasticity. We have found that GPA of planar results are close to GPA of 3D ones. While standard linear elasticity gives strain and rotation fields that are qualitatively different from experiments, discrete periodized planar elasticity agrees qualitatively with them. Quantitative discrepancies and asymmetries may indicate 3D effects and the presence of corrugations, in which case discrete periodized F\"oppl-Von K\'arm\'an equations may improve quantitative agreement with experiments up to effects due to far away defects or to limitations of microscopy or GPA as illustrated in Figure \ref{fig11}. dpPE and dpFvKEs have to be numerically solved to obtain atom positions and, from GPA, strains and rotations. Analytical formulas are much easier to use to fit experimental data. Hyperstress theory is a simple continuum theory that retains second-order lattice effects and therefore may be brought to be close to dpPE by fitting parameters. We have solved it analytically for the elastic fields about a stationary dislocation and fitted its new material moduli by using the experimentally observed strain and rotation. The microscopic dilatation and rotation lengths of hyperstress theory are about 3 and 4 times the graphene lattice spacing. At much larger distances from dislocation points, hyperstress fields become elastic ones. 

Discrete periodized planar elasticity and the discrete F\"oppl-von K\'arm\'an equations are based on  displacement vectors. We have the paradox that, on the one hand, conventional 2D linear elasticity gives strain and rotation fields that differ qualitative and quantitatively from those measured at  dislocation cores in graphene. However 2D linear elasticity gives surprisingly good {\em qualitative} descriptions of the measured strains and rotations in our experiments, provided these fields are calculated by GPA of atomic positions based on the displacement vector of dislocations on a lattice, and not by the usual formulas of continuum mechanics. 

Information about out-of-plane displacements may be obtained from in-plane observations by solving the 3D dpFvKEs and comparing the GPA of numerically calculated configurations and of experiments. In fact, this comparison allows to distinguish between symmetric and antisymmetric configurations of dislocation dipoles and pairs that are possible solutions under zero applied tension. Furthermore, each (symmetric or antisymmetric) configuration of a dislocation dipole or pair admits a mirror configuration with respect to the horizontal plane. In the presence of tension (due to imperfections or far off defects), they can be distinguished and comparison with experimental GPA may be used to decide which theoretical configuration produces the best fit. Our results are clearly applicable to other two-dimensional materials of great current interest \cite{ter12,but13}.

\appendix
\section{Derivation of Eq. (\ref{eq14})}\label{app}
To solve Equation (\ref{eq13}),\cite{ll7} we first extract the part of $\mathbf{u}$ that has a jump discontinuity and is responsible for providing the Burgers vector (\ref{eq12}),
\begin{eqnarray}
&&\mathbf{u}^d=\mathbf{b}\frac{1}{2\pi}\tan^{-1}\frac{y}{x}+(-b_2,b_1)\frac{1}{2\pi}\ln r, \label{eq15}\\
&&\Delta\mathbf{u}^d=(-b_2,b_1)\delta(x)\delta(y),\quad\nabla\cdot\mathbf{u}^d=0, \quad \oint dx_j u^d_{i,j}=b_i.\label{eq16}
\end{eqnarray}
Then the following equation holds for $\mathbf{w}=\mathbf{u}-\mathbf{u}^d$
\begin{eqnarray}
&&(1-l_2^2\Delta)\Delta\mathbf{w}+\left[1+\frac{\lambda}{\mu}+\left(l_2^2-\frac{2l_1^2}{1-\nu}\right)\!\Delta\!\right]\!\nabla(\nabla\cdot\mathbf{w}) 
\nonumber\\
&&\quad =\!\left\{\!2(b_2,-b_1)\!\left[1+\left(l_2^2-\frac{l_1^2}{1-\nu}+\frac{a_1}{\mu}\right)\!\Delta\!\right]\!-\frac{2a_1}{3\mu}(b_2\partial_1-b_1\partial_2)\nabla\right\}\!\delta(x)\delta(y).\label{eq17}
\end{eqnarray}
We use the Green function of the operator $(1-l_2^2\Delta)\Delta$ to write  
\begin{eqnarray}
&&\mathbf{w}=\mathbf{w}^0+\mathbf{w}^1,\label{eq18}\\
&&\mathbf{w}^0=\mathbf{\Gamma}(\nabla)\!\left[\ln r+K_0\!\left(\frac{r}{l_2}\right)\!\right]\!, \label{eq19}\\
&&\mathbf{\Gamma}(\nabla) =\frac{1}{\pi}\!\left\{\!(b_2,-b_1)\!\left[1-\left(\frac{l_1^2}{1-\nu}
-\frac{a_1}{\mu}\right)\!\Delta\!\right]\!-\frac{a_1}{3\mu}(b_2\partial_1-b_1\partial_2)\nabla\right\}\!, \label{eq20}\\
&&(1-l_2^2\Delta)\Delta\mathbf{w}^1+\left[1+\frac{\lambda}{\mu}+\left(l_2^2-\frac{2l_1^2}{1-\nu}\right)\!\Delta\!\right]\!\nabla[\nabla\cdot(\mathbf{w}^0+\mathbf{w}^1)]=0.\label{eq21}
\end{eqnarray}
Clearly $(1-l_2^2\Delta)\Delta(\nabla\times\mathbf{w}^1)=0$. Then $(1-l_2^2\Delta)\nabla\times\mathbf{w}^1$ is a harmonic function that vanishes at infinity. Thus $(1-l_2^2\Delta)\nabla\times\mathbf{w}^1=0$. The Green's identity $\int v(1-l_2^2\Delta)v=\int(v^2+l_2^2|\nabla v|^2)-l_2^2\oint v\partial_nv$ shows that the solution of $(1-l_2^2\Delta)v=0$ that vanishes at infinity is $v=0$. Then $\nabla\times\mathbf{w}^1=0$ and $\mathbf{w}^1=\nabla\phi$. Substituting this into $\sigma_{ij,j}=0$ and using the decay condition at infinity, we obtain the equation
\begin{eqnarray}
\left(2+\frac{\lambda}{\mu}-\frac{2l_1^2}{1-\nu}\Delta\!\right)\!\Delta\phi=-\left[1+\frac{\lambda}{\mu}+\left(l_2^2-\frac{2l_1^2}{1-\nu}\right)\!\Delta\!\right]\!\nabla\cdot\mathbf{w}^0. \label{eq22}
\end{eqnarray}
This equation can be written as 
\begin{eqnarray}
(1-l_1^2\Delta)\Delta\phi &=&\!\left[(1-l_1^2\Delta)-\frac{1-\nu}{2}(1-l_2^2 \Delta) \right]\!\frac{b_1\partial_2-b_2\partial_1}{\pi}\!
\!\left[\ln r\right.\nonumber\\
&+&\left.\!\left(1+\frac{2a_1}{3\mu l_2^2}-\frac{l_1^2/l_2^2}{1-\nu}\right)\!K_0\!\!\left(\frac{r}{l_2}\right)\!\right]\!. \label{eq23}
\end{eqnarray}
Its solution is
\begin{eqnarray}
\phi&=&\frac{b_1\partial_2-b_2\partial_1}{\pi}\!\!\left\{\frac{1+\nu}{8} r^2(\ln r-1)+\!\left(l_2^2+\frac{2a_1}{3\mu}-\frac{l_1^2}{1-\nu}\right)\!\!\left[\ln r+K_0\!\!\left(\frac{r}{l_2}\right)\!\right]\right.\nonumber\\
&+&\left.\!\left(\nu \frac{l_1^2}{2}-\frac{a_1(1-\nu)}{3\mu}\right)\!\!\left[\ln r+K_0\!\!\left(\frac{r}{l_1}\right)\!\right]\!\right\}\!. \label{eq24}
\end{eqnarray}
Since $\mathbf{u}=\mathbf{u}^d+\mathbf{w}^0+\nabla\phi$, putting together (\ref{eq15}), (\ref{eq19}) and (\ref{eq24}), we obtain the displacement vector (\ref{eq14}). We have checked numerically and also using Mathematica that (\ref{eq14}) solves (\ref{eq13}).

\acknowledgments
LLB thanks Amit Acharya for suggesting to look into hyperstress theory. This work has been supported by the Spanish Ministerio de Econom\'\i a y Competitividad grants FIS2011-28838-C02-01,  FIS2011-28838-C02-02, MTM2014-56948-C2-1-P and MTM2014-56948-C2-2-P. LLB and AC thank M.P. Brenner for hospitality during a stay at Harvard University financed by Fundaci\'on Caja Madrid mobility grants. Part of the computations of this work were performed in EOLO, the HPC of Climate Change of the Moncloa International Campus of Excellence, funded by MECD and MICINN.


\begin{thebibliography}{28}
\bibitem{footnote}The singular dislocation point should not coincide with a lattice point.
\bibitem{radii} The radii of the circumferences used to fit the hyperstress elasticity results for a single dislocation are between 0.025 and 0.1 nm. These values are rather small, between 0.1 and 0.4 lattice spacings. Experimental data for larger radii were influenced by far away defects or noise and were not considered.
\end{thebibliography}
\end{document}